\documentclass{pasj00}
\SetRunningHead{Nature of Low-luminosity AGNs}{Tanaka}

\usepackage{times}
\usepackage{ulem}

\begin{document}
\title{A Novel Method to Identify AGNs based on Emission Line Excess and\\the Nature of Low-luminosity AGNs in the Sloan Digital Sky Survey\\II -- Nature of Low-luminosity AGNs}
\author{Masayuki Tanaka}
\affil{Institute for the Physics and Mathematics of the Universe, The University of Tokyo\\  5-1-5 Kashiwanoha, Kashiwa-shi, Chiba 277-8583, Japan}
\email{masayuki.tanaka@ipmu.jp}

\maketitle

\begin{abstract}
We develop a novel method to identify active galactic nuclei (AGNs) and study the nature of
low-luminosity AGNs in the Sloan Digital Sky Survey.
This is the second part of a series of papers and we study the correlations
between the AGN activities and host galaxy properties.
Based on a sample of AGNs identified with the new method developed in Paper-I,
we find that AGNs typically show extinction of $\tau_V=1.2$ and they
exhibit a wide range of ionization levels.  The latter finding motivates us to 
use {\sc [oii]+[oiii]} as an indicator of AGN power.
We find that AGNs are preferentially located in massive, red, early-type galaxies.
By carefully taking into account a selection bias of the Oxygen-excess method,
we show that strong AGNs are located in actively star forming galaxies and
rapidly growing super-massive black holes are located in rapidly growing galaxies,
which clearly shows the co-evolution of super-massive black holes and the host galaxies.
This is a surprising phenomenon given that the growths of black holes and
host galaxies occur at very different physical scales.
Interestingly, the AGN power does not strongly correlate with the host galaxy mass.
It seems that mass works like a 'switch' to activate AGNs.
The absence of AGNs in low-mass galaxies might be due the absence of
super-massive black holes in those galaxies, but
a dedicated observation of nuclear region of nearby low-mass galaxies 
would be necessarily to obtain deeper insights into it.
\end{abstract}

%--------------------------------------------------------------------------
\section{Introduction}

Massive black holes were first speculated as powerhouses for luminous quasars
(e.g., \cite{lynden-bell69}).
A large amount of work conducted afterwards vastly improved our understanding
of active galactic nuclei (AGNs) and it is now widely recognized that 
super-massive black holes are far from rare, exotic objects and they are basic
constituents of massive galaxies.  A large spectroscopic survey of nearby
galaxies has been a successful strategy to unveil statistical nature of AGNs.
\citet{ho95} reported on a systematic survey of 486 nearby galaxies
with the Hale 5m telescope and \citet{ho97a} studied weak AGNs from
the survey, most of which turned out to be low-ionization objects often
called LINERs \citep{heckman80}.
They found that AGNs are preferentially located in massive early-type galaxies
with large bulges.
Further observations of nearby galaxies revealed that super-massive black hole mass correlates
with host galaxy properties, particularly with bulge luminosity and mass
\citep{kormendy95,magorrian98,ferrarese00,gebhardt00}.
This suggests that the super-massive black hole growth and galaxy growth are linked,
and it has motivated a lot of effort to study the black hole evolution
in the context of galaxy evolution both theoretically and observationally.

Recent galaxy formation models, which incorporates energy feedback by AGNs
in one way or another, seem to reproduce observed properties
of galaxies reasonably well.  AGNs seem to be a favorable energy feedback
mechanism due to
their continuous energy output in massive galaxies with QSO-like strong
feedback in starburst phase (e.g., \cite{granato04,springel05,bower06,croton06}),
which helps reproduce the 'break' in the massive end of the stellar
mass function.  However, these simulations cannot directly follow the AGN physics,
which happens on less than a parsec scale, and rely on a simplified recipe.
Also, the exact form of energy feedback is still highly uncertain.
Despite the simplicity and assumptions employed, however, these models
succeeded to reproduce overall properties of observed galaxies relatively well.

Observational evidence of AGN feedback, on the other hand, is still limited
and we do not yet know how exactly AGN affects the galaxy evolution.
High-quality large spectroscopic surveys of the local universe such as 
the 2dF survey \citep{colless03} and the Sloan Digital Sky Survey (SDSS; \cite{york00})
rapidly improved our understanding of AGNs.
It is now widely accepted that AGNs are preferentially located in massive galaxies \citep{kauffmann03}.
Strong AGNs tend to reside in blue galaxies and 
high ionization AGNs (Seyferts) also slightly prefer blue galaxies
over red galaxies \citep{kewley06}.
It seems that very strong AGN activities may be affecting star formation
activities of the host galaxies.
\citet{ho05} studied nearby QSOs and found possibly suppressed SFRs of the host galaxies.
\citet{kim06} extended the analysis with a large number of type-I AGNs drawn
from SDSS and suggested that the observed {\sc [oii]} emission can be
explained entirely by the photo-ionization due to AGN with little emission from HII regions.
Recently, \citet{greene11} examined 15 luminous QSOs and reached essentially
the same conclusion.  
If we turn our attention to low-luminosity AGNs, which comprise the majority
of the overall AGN population, some authors claimed that the host galaxies tend to show colors
in between the red sequence and blue cloud (so-called green valley), which 
led them to speculate on on-going effects of AGN feedback \citep{schawinski07,schawinski10}.
\citet{heckman04} studied the mass of the AGN host galaxies and suggested
that most of the active super-massive black holes are located in intermediate
mass galaxies, therefore representing down-sizing in AGN activities.

A significant fraction of the previous studies from the surveys of the local universe
is based on optical emission line diagnostics first proposed by Baldwin et al. (1981; hereafter BPT)
to identify AGNs thanks largely to a large
number of high quality spectra from SDSS.  However, a major drawback of
the popular BPT diagnostics is that it requires four emission lines
which are not always easy to measure at high signal-to-noise ratios.
Among them, H$\beta$  can be fairly weak in AGNs and that imposes a practical
limit on its sensitivity to low-luminosity AGNs (Paper-I).
%This results in a strong selection bias,
%which has led some of the previous studies to misleading conclusions
%(we will point them out later in the paper).
Several authors proposed variances of the BPT diagnostics to overcome
the issue \citep{rola97,lamareille04,yan06,yan11,juneau11}, but
they may not be fully satisfactory alternatives.
In particular, some of them make a priori assumption
of host galaxy properties and therefore may not be suited to study
AGN - host galaxy correlations.

To improve the situation,
we have developed a novel method to identify AGNs in Paper-I based on 
a very simple idea of comparing expected with observed emission line
luminosities.  We do not
make any a priori assumptions of host galaxy properties.  While the method
suffers from contamination and incompleteness as all the other AGN
identification methods do, it is a statistically efficient method to
identify low-luminosity AGNs.
With this new method, we aim to study the nature of low-luminosity AGNs
and their host galaxy properties in this paper.
%A strong selection bias of BPT was not 
%fully taken into account in most of the previous studies.
%The selection function of our method can be easily characterized and
%we revisit some of the subjects explored by previous studies with the new method.

The structure of this paper is as follows.
In Section 2, we give a brief overview of the new AGN identification method
developed in Paper-I.  We then develop a basis to characterize AGN
activities in Section 3, followed by a detailed study of correlations
between AGN activities and host galaxy properties in Section 4.
We summarize our results and discuss their implications in Section 5.
Unless otherwise stated, we adopt $\Omega_M=0.3$,
$\Omega_\Lambda=0.7$, and $\rm H_0=70\ km\ s^{-1}\ Mpc^{-1}$.
All the magnitudes are given in the AB system.
We use the following abbreviations : AGN for active galactic
nucleus, BPT for the \citet{baldwin81} diagnostics,
SF for star formation and SFR for star formation rate.
Emission lines used in this work include {\sc [oii] $\lambda\lambda3726,3729$},
H$\beta\ \lambda4861$, {\sc [oiii] $\lambda5007$}, {\sc [oi] $\lambda6300$},
H$\alpha\ \lambda6563$, {\sc [nii] $\lambda6583$}, and {\sc [sii] $\lambda6716,6730$}.

%--------------------------------------------------------------------------
\section{Overview of the new method and the data}

\subsection{Method}

We briefly give an overview of the new method to identify AGNs.
For the full development, readers are refereed to Paper-I.

The most widely adopted optical emission line diagnostics to identify AGNs
is the one proposed by \citet{baldwin81} and later extended by
\citet{veilleux87}.  It is based on the idea that an ionizing spectrum
of AGN is often harder than that of young stars, and emission lines
show characteristic intensity ratios.  However, in principle,
a single emission line contains information about AGN.
An emission line luminosity of a galaxy we measure from its
spectrum is

\begin{equation}
L_{measured} = L_{SF} + L_{AGN},
\end{equation}

\noindent
where $L_{SF}$ and $L_{AGN}$ are luminosities due to star formation
and AGN, respectively.
It has been a long standing problem to separate the two components,
which has hindered detailed studies of pure AGN emission.
But, we have developed a novel method to estimate $L_{SF}$ in Paper-I.
AGNs often emit strong emission lines, but their continuum is generally much weaker.
Optical continuum emission of a galaxy spectrum is dominated by
stellar light, not by AGN, in most AGNs \citep{schmitt99}.
The stellar continuum contains information about star 
formation rates (SFRs) and dust extinction of the galaxy.
We extract these two pieces of information by fitting
model templates generated with an updated version of
the \citet{bruzual03} code to the observed spectrum.
From SFRs and dust extinction, we can work out $L_{SF}$.
We then compare $L_{measured}$ with $L_{SF}$.
If we observe a significant $L_{AGN}$, a galaxy likely hosts an AGN.
As we will discuss in Section 3, AGNs exhibit a wide range of
ionization state and we use a sum of {\sc [oii]} and {\sc [oiii]}
luminosities to identify AGNs.
We have shown that we can estimate $L_{SF}$ for {\sc [oii]+[oiii]}
with a factor of $\sim1.7$ accuracy in Paper-I.
Since we identify AGNs using excess Oxygen emission, we dub the method
``Oxygen-excess method''.
Note that we will extensively use stellar mass, SFR and dust
extinction from the spectral fits in this paper.

Compared to the commonly used \citet{baldwin81} diagnostics,
this Oxygen-excess method gives the same SF/AGN classifications
for 85\% of objects that exhibit strong enough emission lines to
apply the BPT method.
The BPT diagnostics uses 4 emission lines and it takes ratios of them,
requiring a high signal-to-noise ratio for each line.
On the other hand, the new method uses a sum of the two lines and
its requirement for signal-to-noise is less demanding.
As a result, the method is applicable to 78\% of the objects
in our sample, while BPT is applicable only to 43\%.
Our method is thus fairly sensitive to low-luminosity AGNs.
We have quantified average properties of the Oxygen-excess objects
by stacking the spectra and we have shown that many of
the Oxygen-excess objects, if not all, likely host AGNs.
Serendipitous X-ray observations \citep{evans10}
as well as radio data from the FIRST survey \citep{becker95,white97}
also show that SF/AGN classifications are made well.
Another unique feature of the method is that it allows us to
subtract emission line luminosities due to star formation and
extract AGN luminosities, which are crucial to characterize AGNs.
We will make a full use of these unique features to study
the nature of low-luminosity AGNs in this paper.

\subsection{Data}

We use data from the Sloan Digital Sky Survey Data Release 7 \citep{abazajian09}.
Using a dedicated 2.5m telescope \citep{gunn06}, the SDSS surveyed
a quarter of the sky in both imaging and spectroscopy 
\citep{fukugita96,gunn98,eisenstein01,strauss02,richards02,gunn06,doi10}.
The galaxy sample for our study is drawn from the main galaxy
sample \citep{strauss02}, which is a flux-limited sample
down to $r=17.77$ selected from the imaging survey.
We apply the following selection criteria:
{\sc specClass=2} (i.e., objects are galaxies) located at
$0.02<z<0.10$  with high confidence flags ({\sc zConf$>$0.8}
and {\sc zWarning=0}).
We intentionally remove QSO-like objects ({\sc specClass=3})
because our method is not applicable to those objects (continuum
needs to be dominated by stars, not by AGN), leaving 283,031
objects in total.
We apply a first-order
correction for the fiber loss to the stellar mass and SFR
by assuming that light in the fiber is representative of
the entire galaxy.  The corrected quantities are shown with
a subscript {\it apercorr}.

\subsection{Caveats on photo-ionization due to non-AGN sources}
Before we present our results, it is important to remind of the contamination
of objects that are photo-ionized by non-AGN sources.
As discussed in Paper-I, there are a number of possible
non-AGN sources that could produce LINER-like emission line ratios.
They include shocks \citep{cox72,heckman80,dopita95,dopita96},
Wolf-Rayet stars \citep{terlevich85,kewley01},
post-starburst \citep{taniguchi00}
and post-AGB stars \citep{binette94,stasinska08,sarzi10,cidfernandes11}.
Relative contributions of these ionizing sources to the observed
LINER populations are uncertain, but they may contaminate
our AGN sample.

As shown later, most of the Oxygen-excess objects are red, massive galaxies.
In these galaxies, shocks, Wolf-Rayet stars, and post-starburst
are unlikely the primary origin of the observed emission.
Post-AGB stars remain a concern.  Several attempts have been made
to quantify their role \citep{binette94,stasinska08,sarzi10,cidfernandes11,yan11b}.
In particular, \citet{sarzi10} and \citet{yan11b} studied nearby galaxies and
observed spatially extended H$\alpha$ or H$\beta$ emission, which is unlikely to be fully
explained by a central source. 
There is no doubt that these post-AGB stars contaminate the true AGN population
and the observed emission line luminosities of the Oxygen-excess
objects are at least partly contaminated by post-AGB photo-ionization.

We argue in Section 3.3 that emission due to post-AGB stars does
not seem to dominate the overall emission, but the exaction fraction of
objects that are photo-ionized by post-AGB stars among the Oxygen-excess
objects and the exact contribution of
post-AGB stars to the observed emission remain unclear.
As discussed in Paper-I, the SDSS data are not a right data set to
precisely pin down the role of post-AGB stars and we do not try to
pursue the issue further in this paper.
But, we should bear in mind that this unknown role of post-AGB stars
remains one of the major uncertainties throughout the paper.

%--------------------------------------------------------------------------
\section{Dust in AGNs and AGN power indicator}

First of all, we develop a basis to characterize AGN activities.
We first quantify dust extinction in AGNs.  We then study emission line properties
using the extinction corrected luminosities and examine indicators of the AGN power.
Finally, we revisit the issue of measuring SFRs of
the host galaxies from emission lines.

%------------------------------------------
\subsection{Extinction in AGNs}

%---------------------
\begin{figure}
  \begin{center}
    \FigureFile(80mm,80mm){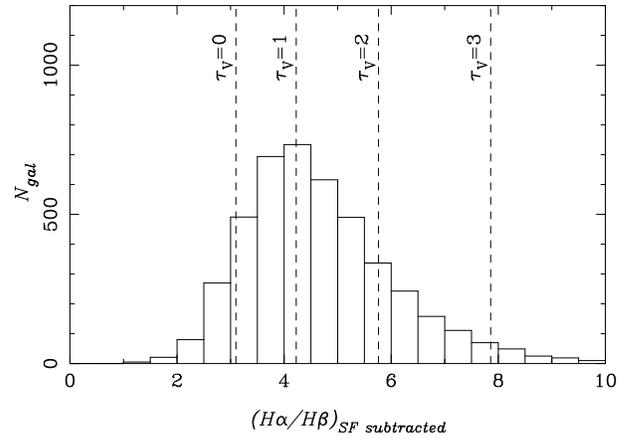}
  \end{center}
  \caption{
    Distribution of the Balmer decrement.
    The star formation component is subtracted from
    the H$\alpha$ and H$\beta$ luminosities
    and extinction due to the interstellar medium is corrected for.
    The dashed lines show $\tau_V$ assuming the dust-free
    Balmer decrement of $3.1$.  Note that one can convert
    $\tau_V$ into the commonly used extinction in the $V$-band using $A_V=1.09\tau_V$.    
    We use objects with H$\alpha$ and H$\beta$ detected at more than
    $10\sigma$ only here to quantify the line ratio.
  }
  \label{fig:agn_balmer_decr}
\end{figure}
%---------------------

AGNs can be obscured by dust around the nuclear regions
and it is essential to correct for the dust extinction in order to quantify
AGN activities.  We measure emission line luminosities due to
AGNs in a two-step procedure.  First, we subtract the star formation
component, $L_{SF}$, from a measured emission line luminosity.
We then correct for the dust extinction
% due to the ambient interstellar medium of the galaxies
to obtain emission line luminosities coming out of the central region.
To be specific, in case of H$\alpha$,

\begin{equation}
L_{H\alpha}=(L_{H\alpha,meas} - L_{H\alpha,sf})\times \exp(0.3\times0.75\times\tau_V),
\end{equation}

\noindent
where $L_{H\alpha,meas}$, $L_{H\alpha,sf}$ and $\tau_V$ are the measured
H$\alpha$ luminosity, 'predicted' H$\alpha$ luminosity due to
star formation, and the optical depth in the $V$-band from
the spectral fitting.  The factor of 0.3 to $\tau_V$ is due to
the assumption that 30\% of the dust extinction comes from the ambient
interstellar medium \citep{charlot00}.  The term $0.75\tau_V$ is the optical depth at
the wavelength of H$\alpha$ assuming the \citet{cardelli89} extinction curve.
We note that we tend to underestimate $\tau_V$ from the spectral
fitting (see Fig. 1 of Paper-I) and the extinction due to the interstellar
medium may not be fully corrected\footnote{
We underestimate $\tau_V$ systematically by $\Delta \tau_V\sim1$ for most objects
as shown in Fig.1 of Paper-I.
If we empirically correct for the offset, we obtain a median AGN extinction of
$\tau_{V}=0.7$, which is smaller than that obtained below ($\tau_{V}=1.2$).
}.

We apply the same equation to $L_{H\beta}$ and plot the Balmer decrement
of the AGN emissions in Fig. \ref{fig:agn_balmer_decr}.
Here we adopt the dust-free Balmer decrement of 3.1 for AGNs
\citep{halpern83,ferland83,gaskell84}.
We also have confirmed that there is no change in our main
conclusions if we adopt a decrement of 2.86 (case 'B' recombination for
$T=10^4\ \rm K$ and $N_e=10^2\ \rm cm^{-3}$; \cite{osterbrock06}).
The Oxygen-excess objects typically have $\tau_{V}=0-2$ extinction
with the median being $\tau_{V}=1.2$. 
We note that the median value does not change if we include
lower significance detections of the lines (e.g., $3\sigma$),
although the scatter significantly increases.
\citet{dahari88} showed that Seyfert 2's typically show
the Balmer decrement of $3-6$, which is fully consistent with our estimates here.
\citet{kauffmann03} quoted a median extinction of $A_V\sim3$ mag.
in AGNs using data from SDSS.  Our extinction is less than that, but
their large value is due to the assumption
of the relatively flat extinction curve of \citet{charlot00}.
We have confirmed that we obtain a similar amount of extinction 
if we use the \citet{charlot00} curve.
It is interesting to note that the observed range of Balmer decrement is similar
to that observed in Seyfert galaxies \citep{osterbrock77,greene05}.

For the Balmer decrement measurements, we need good signal-to-noise
ratios on the Balmer lines, but H$\beta$ can often be noisy
(and that is the reason why the BPT diagnostics is not applicable
for a large number of galaxies).
It is difficult to measure the Balmer decrement for individual galaxies
and we therefore apply a constant extinction correction
regardless of the properties of the AGNs and host galaxies.
We note that we do not observe strong dependence of
the balmer decrement on AGN power and thus the constant correction
is a reasonable first order correction.
The total optical depth of Oxygen-excess objects is

\begin{equation}
\tau_{V,total}=0.3\times\tau_{V}+1.2.
\end{equation}

\noindent
The first term is the extinction due to the ambient interstellar
medium, which is derived from the spectral fitting.
The second term is the extinction around the nuclear region.
To be specific, we replace $0.3\times\tau_V$ in Eq. 2 with Eq. 3
to obtain intrinsic line luminosities due to AGN.
We use a subscript {\it AGN} when we
show star formation subtracted, extinction corrected emission line luminosities
of the Oxygen-excess objects.  This is just for simplicity and
we do not mean that all of the Oxygen-excess objects are real AGNs.

%------------------------------------------
\subsection{A new indicator of AGN power and comparisons with X-ray power}

%---------------------
\begin{figure}
  \begin{center}
    \FigureFile(80mm,80mm){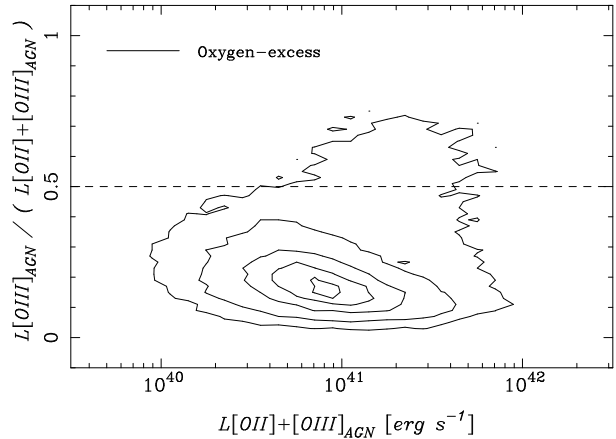}\\
  \end{center}
  \caption{
    Fraction of {\sc [oiii]$_{AGN}$} luminosity to {\sc [oii]$_{AGN}$+[oiii]$_{AGN}$} luminosity
    plotted against {\sc [oii]$_{AGN}$+[oiii]$_{AGN}$}.
    The solid contours are Oxygen-excess objects and their star formation
    fluxes are subtracted off and the extinction is corrected for using Eq. 3.
    The horizontal dashed line shows the equality of {\sc [oii]$_{AGN}$} and {\sc [oiii]$_{AGN}$}
    luminosities.  {\sc [oii]$_{AGN}$}$<${\sc [oiii]$_{AGN}$} holds above the line, and
    {\sc [oii]$_{AGN}$}$>${\sc [oiii]$_{AGN}$} holds below the line.
  }
  \label{fig:oiioiii_frac}
\end{figure}
%---------------------

Having quantified extinction in AGNs, we are now ready to characterize
AGN activity.
In order to quantify AGN power, we ideally want to integrate
all the photons emitted from the central engine over the whole wavelength range.
But, this is observationally very challenging to achieve.
A luminosity of the {\sc [oiii]} line is often
used as an indicator of AGN power and a clear correlation between
the absorbed {\sc [oiii]} luminosity and hard X-ray luminosity has been observed
in a hard X-ray selected sample \citep{heckman05}.
Recently, \citet{degasperin11} also presented a tight correlation between
absorbed hard X-ray luminosity and {\sc [oiii]} luminosity.
The use of {\sc [oiii]} as a proxy for AGN power is partly motivated by
a commonly adopted assumption that
contamination from star formation is relatively small in {\sc [oiii]}.
But, {\sc [oiii]} is a relatively high ionization line.  Is it
a good indicator of AGN power for low-luminosity AGNs studied here?

If we look at ionization state of the Oxygen-excess galaxies in Fig.
\ref{fig:oiioiii_frac}, we find that AGNs spans a wide range in ionization state.
\citet{heckman80} first introduced a class of low-ionization emission
line galaxies (LINERs).  The original definition of LINERs is
{\sc [OII]}$>${\sc [OIII]} and {\sc [OI]}$>0.33${\sc [OIII]}.
\citet{ho93} showed that this definition is essentially the same as {\sc [OIII]}/H$\beta>3$.
Due to the weak H$\beta$ emission in most of the Oxygen-excess galaxies,
we further revise the definition to {\sc [OII]$_{AGN}$}$>${\sc [OIII]$_{AGN}$}, which is
equivalent to {\sc [OIII]}/H$\beta>3$ due to the tight correlation between
{\sc [OII]/[OIII]} and {\sc [OIII]}/H$\beta$ \citep{baldwin81}.
Note that we use star formation subtracted, extinction corrected luminosities
for the LINER/Seyfert classification.
At low luminosities, AGNs are mostly LINERs.
Seryfert-like galaxies with {\sc [OII]$_{AGN}$}$<${\sc [OIII]$_{AGN}$} appear
at high luminosities.
We find that 7\% of the Oxygen-excess objects have {\sc [oii]$_{AGN}$}$<${\sc [oiii]$_{AGN}$}.
This fraction should not be taken as a general fraction of Seyferts
among AGNs because we intentionally excluded QSO-like objects from the sample.

{\sc [oiii]} is often used as an indicator of AGN power
\citep{heckman05}, but given the relatively wide range of ionization states of
the Oxygen-excess objects, we deem that the sum of
{\sc [oii]+[oiii]} is a better indicator of the AGN power.
In general, {\sc [oii]} and {\sc [oiii]} lines balance
each other in gaseous nebulae. {\sc [oiii]} is strong in high ionization states
and {\sc [oii]} is strong in low ionization states.
A sum of the two lines is more robust to variations in ionization
levels than either one of them and is a better indicator of total energy
emitted by Oxygen.
H$\alpha$ line would be another straightforward option.
However, as shown in Paper-I, AGNs exhibit relatively weak
H$\alpha$ and
the AGN component does not stand out.  If we subtract the star
formation component as we do for {\sc [oii]} and {\sc [oiii]},
we obtain a negative H$\alpha$ luminosity for a number
of Oxygen-excess objects.  This means that our SFR and $\tau_V$ estimates from the spectral
fits are not sufficiently good to extract the relatively weak
H$\alpha$ due to AGN.  On the other hand, the collisionally
excited lines are normally strong in AGNs.
For these reasons, we choose {\sc [oii]+[oiii]} to identify
AGNs and to characterize them.  If one could improve the estimates of
H$\alpha$ flux due to star formation, H$\alpha$ would be
a better indicator because it is less sensitive to ionization,
dust, and metallicity.

We now test this new indicator of the AGN power by comparing with
hard X-ray luminosities, which is likely a very good indicator of
the AGN power \citep{ho08}.
As described in Paper-I, we use the Chandra-SDSS matched catalog
release 1.1 \citep{evans10}.
We require the sources to be point-like and located within
2 arcsec from the optical center to ensure that X-rays
likely originate from the center of the galaxies.
We compare the optical emission line luminosities with 
X-ray luminosities to verify the effectiveness of
the new AGN power indicator.  We note that we remove X-ray
sources that may be non-AGN origin from the comparisons
(Fig. 11 of Paper-I show how we identify those sources). 
We further note that we do not correct for absorption in hard X-rays.
The typical Galactic neutral hydrogen column density in our sample is
low (a few times $10^{20}\rm cm^{-2}$; \cite{dickey90}) and intrinsic
absorption is likely below  $10^{22}\rm cm^{-2}$ within the luminosity
range we explore here \citep{mainieri07}.
At this column density, absorption in the hard band is small \citep{morrison83}.
Our hard X-ray luminosity should therefore be a reasonable estimate.

%---------------------
\begin{figure*}
  \begin{center}
    \FigureFile(52mm,120mm){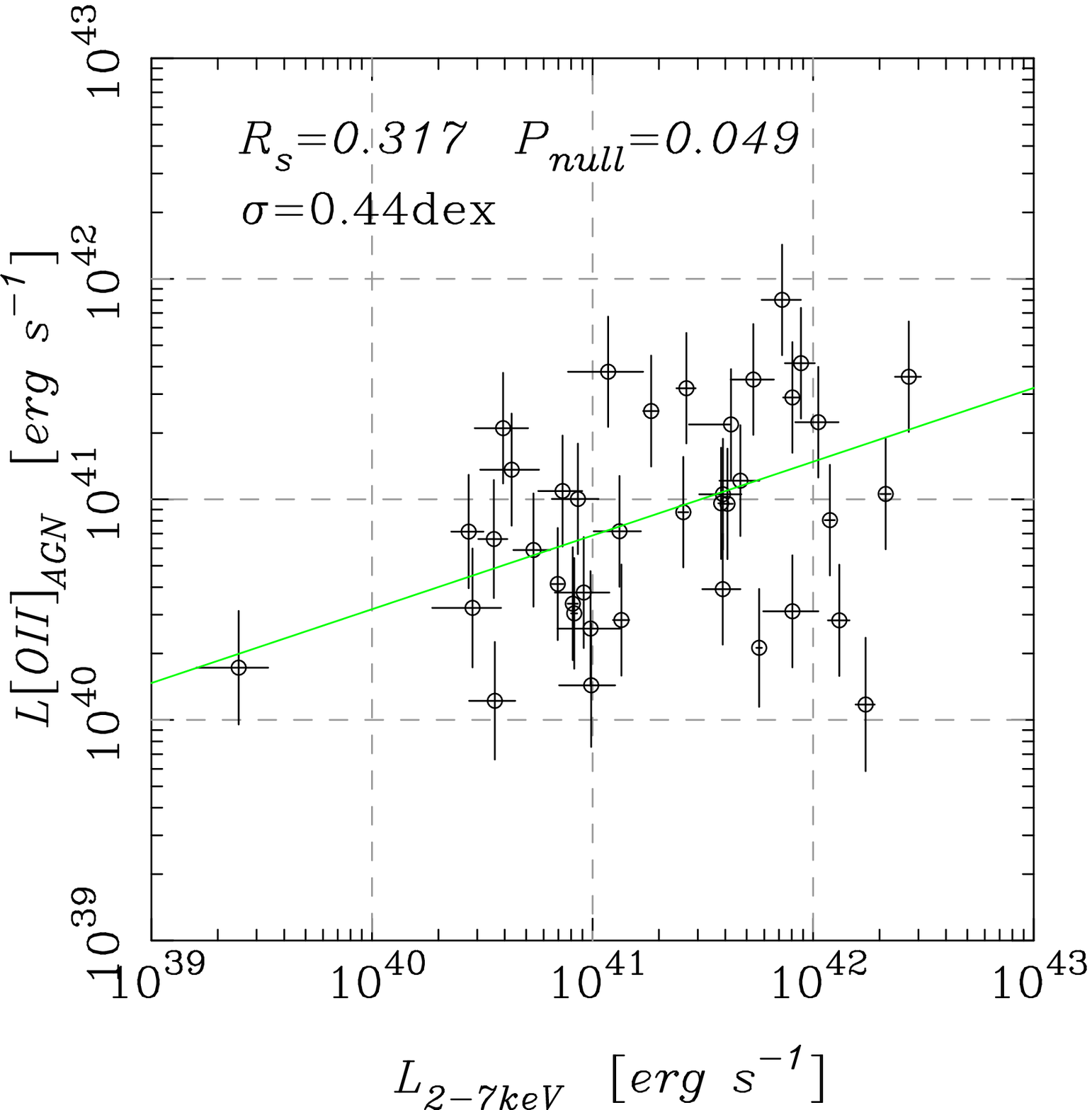}\hspace{0.5cm}
    \FigureFile(52mm,120mm){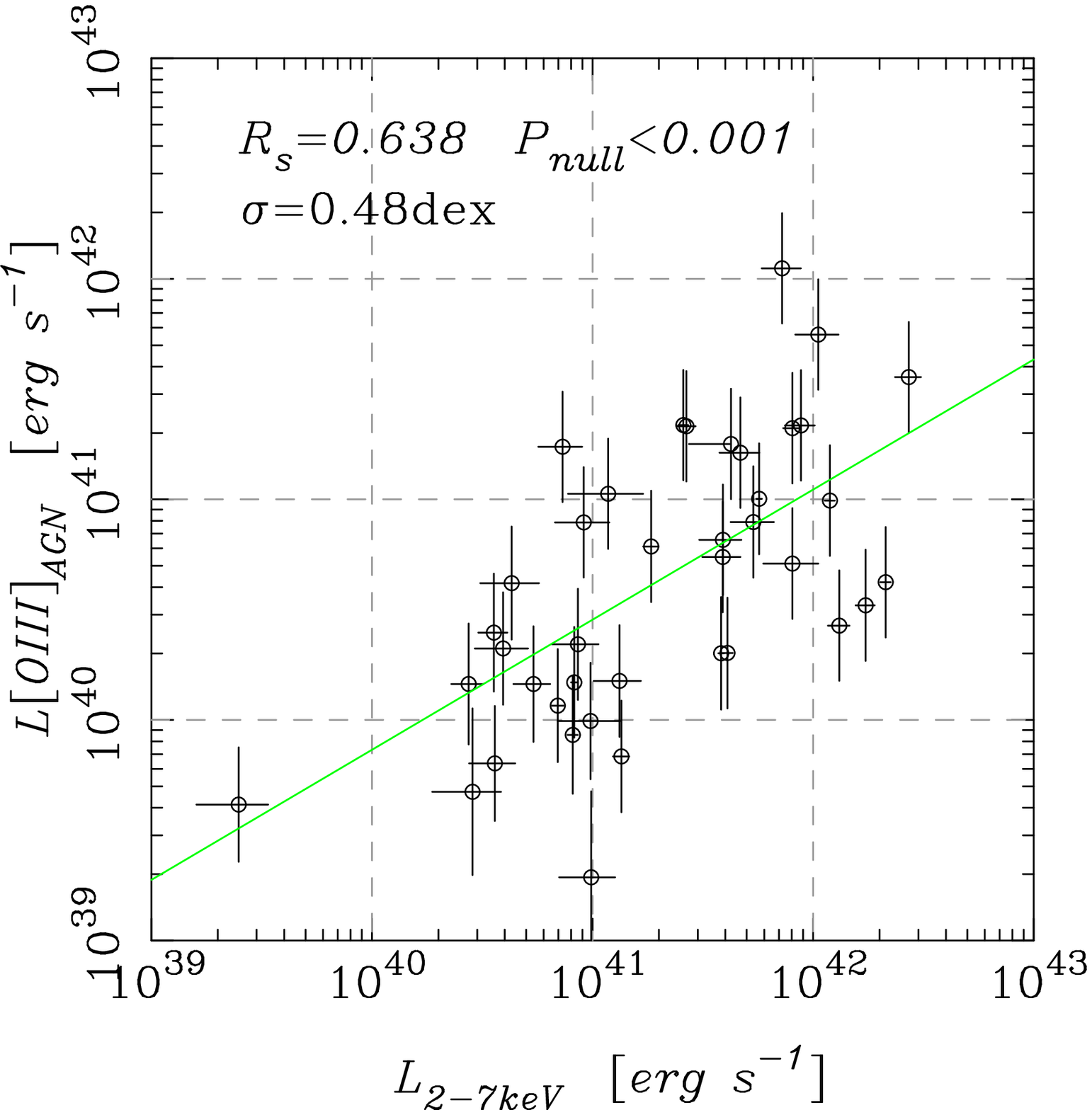}\hspace{0.5cm}
    \FigureFile(52mm,120mm){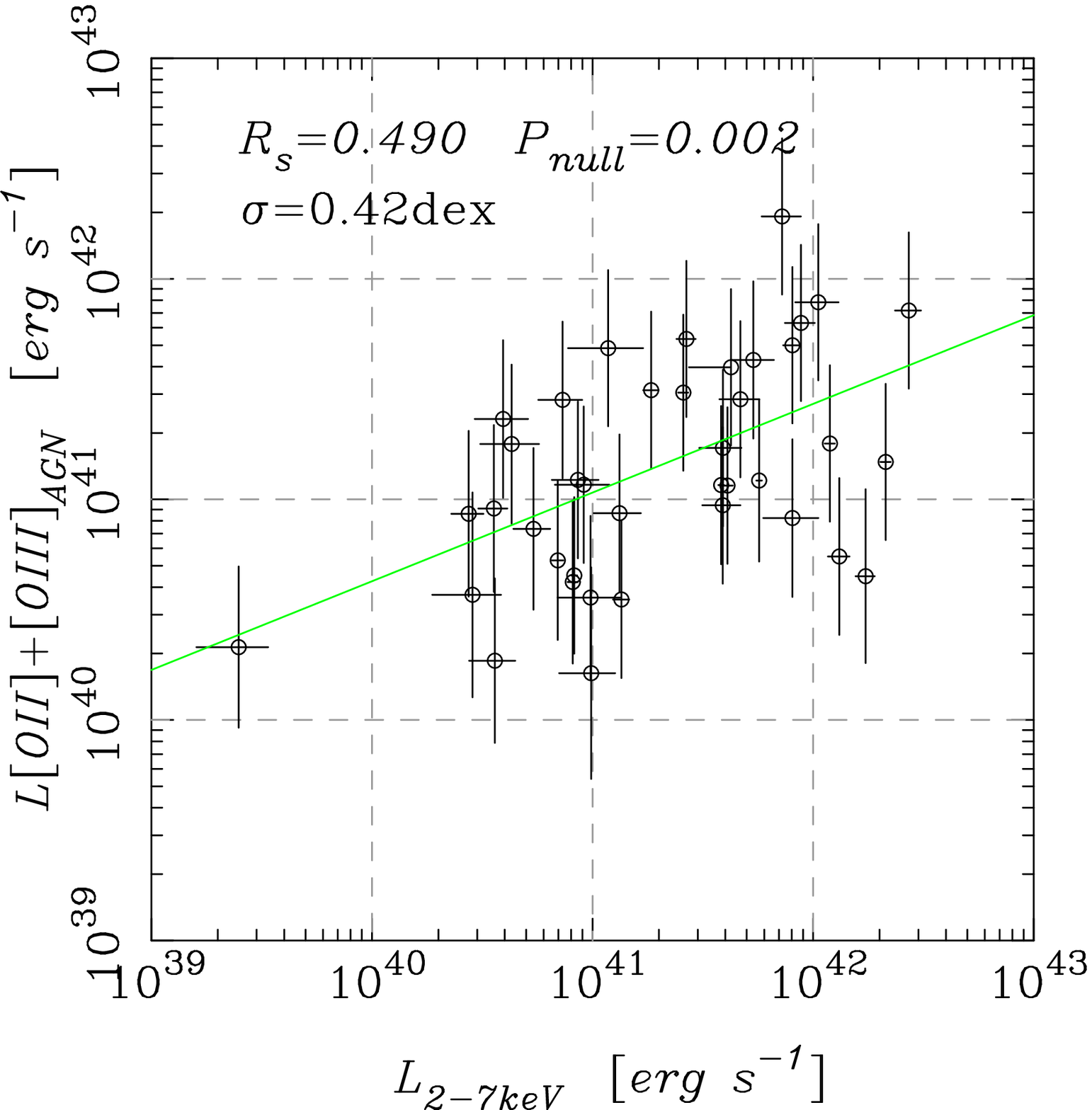}\\
  \end{center}
  \caption{
    Optical emission line luminosities plotted against hard X-ray luminosity.
    The plots show {\sc [oii]$_{AGN}$}, {\sc [oiii]$_{AGN}$}, and {\sc [oii]+[oiii]$_{AGN}$}
    from left to right.
    We remove soft low-$L_X$ sources, some of which may not be AGNs, from the plots.
    The numbers in each plot show a Spearman's rank correlation
    coefficient and a null probability.
    The solid lines are biweight fits to the data points and the dispersion
    in the optical luminosity around them are also shown in the plots.
  }
  \label{fig:oiioiii_xray}
\end{figure*}
%---------------------

We show comparisons between hard X-ray luminosity and optical emission
line luminosities in Fig. \ref{fig:oiioiii_xray}.  
We show not only {\sc [oii]+[oiii]$_{AGN}$}, but {\sc [oii]$_{AGN}$} and
{\sc [oiii]$_{AGN}$} as well. 
We find that all the lines show
statistically significant correlations
with X-rays.  We perform the Spearman's rank test and find the positive
correlation with a high significance ($>2\sigma$) to reject the null hypothesis
in all the lines.  Also, we perform the biweight fits to the data points
and show the best-fitting lines in the plots.
The dominant source of error in our AGN flux estimates is in our
prediction of emission line fluxes due to star formation, which is 
0.23 and 0.29 dex for {\sc [oii]} and {\sc [oiii]}, respectively.
We adopt it as an uncertainty in the subtraction of
the star formation component for each line.  The biweight fits are performed
taking these uncertainties into account.
We find that {\sc [oiii]$_{AGN}$} shows a stronger correlation with the X-ray power than
{\sc [oii]$_{AGN}$} \citep{simpson98}.
Contrary to our expectations, {\sc [oii]+[oiii]$_{AGN}$} shows somewhat
poorer correlation than {\sc [oiii]$_{AGN}$} alone, although the scatter decreases slightly.
Apparently, the relatively poor correlation between {\sc [oii]$_{AGN}$} and X-ray
luminosities weaken the correlation in {\sc [oii]+[oiii]$_{AGN}$}.
We suspect that it might be due to insufficient extinction correction.
Due to our limited ability to measure the extinction in individual galaxies,
we apply the constant correction for dust due to AGN to all the galaxies.
But, this is probably just a rough correction and a small over/under
correction introduces a larger error in {\sc [oii]} than in {\sc [oiii]}
because {\sc [oii]} is at a shorter wavelength.

Now, we face a difficult question of the choice of the AGN power indicator.
We would like to use {\sc [oii]+[oiii]} to keep consistency with the
identification method.  As shown in Fig. 7 of Paper-I, the stacked
spectra of the Oxygen-excess objects show stronger {\sc [oii]} than
{\sc [oiii]} and {\sc [oii]} is a very powerful feature to identify
Oxygen-excess objects.  On the other hand, we would like to use
{\sc [oiii]} only as a better indicator of the AGN power at the cost of
loosing consistency with the identification process somewhat.
We have performed all the analyses presented in the paper using
the two AGN power indicators and found that our conclusions remain
totally the same regardless of the choice of the indicator.
Given this robustness of our conclusions, we choose {\sc [oii]+[oiii]}
as an indicator of the AGN power to keep the consistency with
the AGN identification scheme.

%---------------------
\begin{figure}
  \begin{center}
    \FigureFile(80mm,120mm){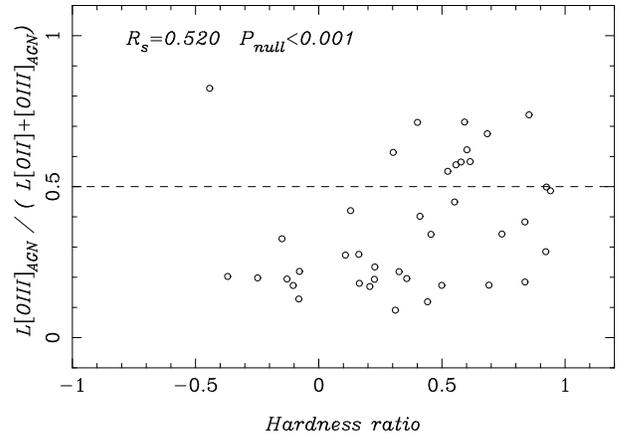}\\
  \end{center}
  \caption{
    {\sc [oiii]$_{AGN}$}/({\sc [oii]$_{AGN}$+[oiii]$_{AGN}$}) plotted
    against X-ray hardness ratio between soft (0.5-1.2 keV) and hard
    (2-7keV) bands.
    The numbers show a Spearman's rank correlation coefficient and
    a null probability.
  }
  \label{fig:oiioiii_xray2}
\end{figure}
%---------------------

We note in passing that we find an interesting correlation
between ``optical hardness'' and X-ray hardness in Fig.
\ref{fig:oiioiii_xray2}.  The optical line ratio plotted is
essentially a simple function of {\sc [oii]/[oiii]}, which is
tightly correlated with {\sc [oiii]/}H$\beta$ and is a good indicator
of the ionization level \citep{baldwin81}.
The Spearman's rank correlation coefficient suggests a
positive correlation between the optical and X-ray hardness
with a statistically significant probability to reject
the null hypothesis.
This correlation can be understood as a product of two other correlations:
one between $L_{X}$ and hardness, and the other between emission line luminosity
(which is correlated with $L_{X}$) and ionization state.
We find that most of the hard sources in Fig. \ref{fig:oiioiii_xray2}
are powerful AGN with $L_X>10^{41.5}\rm\ erg\ s^{-1}$.
The AGN power then correlates with the ionization state as we will show in Section 4.
These two correlations result in the trend in Fig. \ref{fig:oiioiii_xray2}.

%------------------------------------------
\subsection{Possible evidence against a dominant role of post-AGB photo-ionization}

Let us  go back to the issue of objects with non-AGN photo-ionization.
As we discussed in Paper-I and earlier in this paper,
photo-ionization due to post-AGB stars can be important in LINERs.
This comes from a few observational pieces of evidence.
One is an energy budget argument.  AGNs in LINERs tend to show
a deficit in the amount of ionizing photons to fully explain
the observed emission line luminosity \citep{maoz98,eracleous10b},
requiring an additional source of ionization mechanism
other than AGN.
Another piece of evidence is that spatially extended emission is observed
in LINERs using 2D spectrograph \citep{sarzi10}.
Recently, \citet{yan11b} also detected extended emission by
a clever analysis of extended emission using varying redshift slices.
We suspect that they suffer from contamination of star forming galaxies
because a cut in $D_{4000}$ is not enough to eliminate them from their sample
as we mention below. 
Extended emission does not necessarily
support the post-AGB origin of the emission \citep{yan11b}, but spatially
extended post-AGB stars remain a natural explanation of it.
It is not clear whether post-AGB stars alone can produce enough 
ionizing photons to fully explain the observed emission
(e.g., \cite{yan11b} reports on the deficit of ionizing photons even
if they assume that 100\% of the ionizing photons are absorbed to
produce emission line, while \cite{sarzi10} claims that they have
enough ionizing photons).
But, there is no doubt that emission due to post-AGB stars contaminate
the emission line luminosities observed in the Oxygen-excess objects.
The question is how much post-AGB stars contribute to the overall emission.
To address this question, we focus on the correlation between optical
emission line luminosities and hard X-ray luminosities in Fig. \ref{fig:oiioiii_xray}.

\begin{figure}
  \begin{center}
    \FigureFile(60mm,80mm){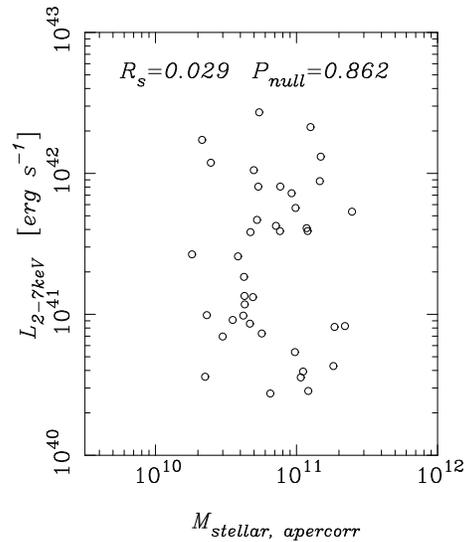}
  \end{center}
  \caption{
    Hard X-ray luminosity plotted against stellar mass.
    Objects that are likely contaminated by non-AGN sources
    (those in the dashed box in Fig. 11 of Paper-I)
    are removed.  We also show the Spearman's rank correlation
    coefficient and null probability.
  }
  \label{fig:oiioiii_xray5}
\end{figure}

Post-AGB stars are not luminous in hard X-rays and
the observed hard X-ray luminosity is likely primarily due to AGNs.
One may worry that low-mass X-ray binaries may contribute to the hard X-ray
luminosity.  But, we show in Fig. \ref{fig:oiioiii_xray5} that
the hard X-ray luminosity does not correlate with host galaxy stellar mass.
The Spearman's rank correlation coefficient is very small and the test does not
reject the null hypothesis.  This rather weak (or no) dependence of
X-ray luminosity and host galaxy mass has also been noted by other authors
\citep{mullaney11,aird11}.
This strongly rules out the low-mass X-ray binary origin of the hard X-ray \citep{kim04}.
The observed hard X-ray luminosity is primarily due to AGNs.

%---------------------
%---------------------

On the other hand, optical emission can be contaminated by post-AGB photo-ionization.
The correlation between optical emission line luminosity and hard X-ray luminosity
can therefore be used to gauge the contamination from post-AGB stars.
The observed clear correlation between them shown in Fig. \ref{fig:oiioiii_xray}
suggests that the post-AGB contamination is not significant.
If the contamination was significant, we would not have observed such a 
clear correlation between hard X-ray and, e.g., {\sc [oii]+[oiii]} luminosities.
We shall note that the X-ray sources are drawn from 
serendipitous Chandra observations and they have typical properties of
the Oxygen-excess objects, i.e., they are massive galaxies with
$>10^{10}\rm M_\odot$ in stellar mass and are red quiescent galaxies
with SFRs below $0.1\rm M_\odot\ yr^{-1}$.
Therefore, these X-ray sources should be representative of the Oxygen-excess
objects.

The X-ray sample is a small subset of the entire Oxygen-excess objects.
To further quantify the emission due post-AGB stars, we define a subsample
of the Oxygen-excess objects located at $0.055<z<0.060$ and with zero SFRs
from the spectral fits.  The narrow redshift slice is adopted to minimize
the variation of our sensitivity to emission line luminosities within
the redshift slice.
  The redshift range is arbitrarily chosen, but
our conclusion below does not strongly depend on a particular choice of the redshift range.
The zero SFR is to ensure that the host galaxies
are not forming stars so that we do not have to correct for emission
line luminosities due to star formation.
The observed emission is therefore due to AGN and to post-AGB star photo-ionization
(if present) in this subsample.

%---------------------
\begin{figure}
  \begin{center}
    \FigureFile(60mm,80mm){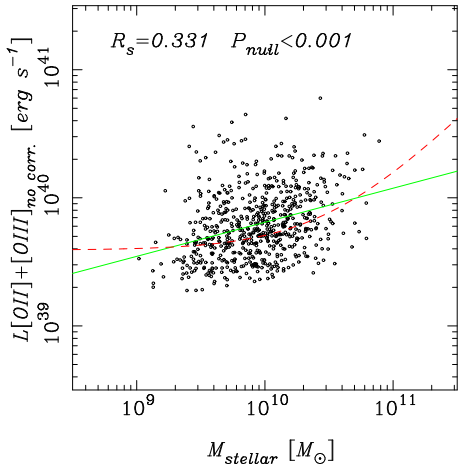}
  \end{center}
  \caption{
    {\sc [oii]+[oiii]} luminosity plotted against stellar mass.
    The Oxygen-excess objects plotted here are located at $0.055<z<0.060$
    and are not forming stars.
    Extinction defined in Eq. 3 is not applied and thus the luminosity is a 'raw' luminosity.
    Stellar mass is not corrected for the aperture loss.
    The solid line is a biweight fit to the data with a power-law slope of 0.26.
    The dashed curve shows the best fit model to the data (see text for details).
  }
  \label{fig:bhm_corr21}
\end{figure}
%---------------------

We show {\sc [oii]+[oiii]} luminosity against stellar mass of this subsample
in Fig. \ref{fig:bhm_corr21}.
Here, we do not apply any dust
correction or aperture correction, i.e., 'raw' emission line luminosity
within the fiber aperture is plotted against stellar mass within the fiber.
The emission due to post-AGB stars should correlate strongly with
stellar mass contained within the same area covered by the fibers.
On the other hand, the AGN emission is unlikely to be strongly correlated
with mass.  Therefore, the correlation in Fig. \ref{fig:bhm_corr21}
is a rough measure of the contribution of the post-AGB stars.

The Spearman's rank test suggests that there is a correlation between
the emission line luminosity and stellar mass.  The null hypothesis is
rejected at a significant level.  This suggests that the post-AGB stars
do contribute to the observed line luminosities.  However, the observed
correlation is relatively weak -- the biweight fit gives a log-linear
slope of 0.26.  That is, if stellar mass increases by an order of magnitude,
the emission line luminosity increases by less than a factor of 2.
The abundance of post-AGB stars should depend on star formation histories
of galaxies, which then depend on galaxy mass.  However, we are looking
only at massive quiescent galaxies\footnote{Stellar mass plotted in Fig.
\ref{fig:bhm_corr21} is not corrected for the aperture loss and is
significantly smaller than a typical stellar mass of the Oxygen-excess
objects discussed later in this paper, which is corrected for the aperture loss.},
and it is very unlikely that
their star formation histories vary so widely that the stellar mass
dependence of the post-AGB abundance is smoothed out.
The most naive interpretation of Fig. \ref{fig:bhm_corr21}  would be that
post-AGB stars do contribute to the observed line emission, but 
their contribution is minor.

Let us make an attempt to quantify the contribution of post-AGB stars
using a very simple model.
If we assume that emission due to post-AGB stars linearly increase
with stellar mass and emission due to AGN does not depend on stellar mass, 
we can write

\begin{equation}
L_{obs}=L_{AGN}+\alpha\times M_{stellar},
\end{equation}

\noindent
where $L_{obs}$, $L_{AGN}$, $\alpha$ are observed luminosity,
luminosity due to AGN, and luminosity due to post-AGB photo-ionization
per unit stellar mass.
Let us modify the second term to define post-AGB emission
relative to $L_{AGN}$ at $M_{stellar}=10^{10}\rm M_\odot$. 

\begin{equation}
L_{obs}=L_{AGN} \left(1+\beta\times\frac{M_{stellar}}{10^{10}\rm M_\odot}\right).
\end{equation}

\noindent
If $\beta=1$, AGN and post-AGB equally contribute to the observed emission
at $M_{stellar}=10^{10}\rm M_\odot$.
We allow $L_{AGN}$ and $\beta$ to vary and fit the observed data in Fig. \ref{fig:bhm_corr21}
using the $\chi^2$ statistics. We find $\log L_{AGN}=39.707\pm0.005$
and $\beta=0.305\pm0.011$ give the best fit.
The errors are derived by taking $\Delta \chi^2=1$ and are very small,
but we should note that we ignore intrinsic scatter in $L_{AGN}$.
The best-fit model is shown as the dashed curve in the figure.
The fit suggests that $23\%$ of the observed emission is due to post-AGB stars
at $M_{stellar}=10^{10}\rm M_\odot$.  Because of the assumptions we made,
this is nothing more than a rough guess.
But, it seems that post-AGB stars do not play a dominant role.

%---------------------
\begin{figure}
  \begin{center}
    \FigureFile(80mm,80mm){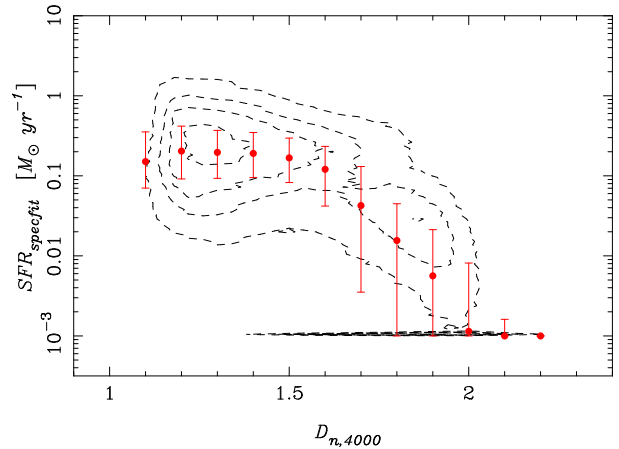}
  \end{center}
  \caption{
    SFR plotted against $D_{n,4000}$.
    The contours are all the objects in our parent sample (i.e., Main galaxies at $0.02<z<0.10$).
    The points and error bars show the median and quartile of the SFR distribution.
  }
  \label{fig:d4000_sfr}
\end{figure}
%---------------------

This result may not appear consistent with those from the literature
(e.g., \cite{sarzi10,yan11b}).
We do not try to resolve the issue as the SDSS data is unlikely a right
data set to address it as we mentioned in Paper-I, 
but we shall recall that our concern is
in a very low luminosity regime.  We have to be sure that we do not
suffer from low-level star formation that could happen in early-type galaxies.
To illustrate its importance, we show SFR against $D_{n,4000}$ in Fig. \ref{fig:d4000_sfr}.
$D_{n,4000}$ is a sensitive index to age of the stellar population and
is used in earlier work to select quiescent galaxies.
However, it is hard to construct a pure sample of quiescent galaxies with
this index alone.  Even if galaxies have a large $D_{n,4000}$,
they may well be undergoing low-level star formation as shown in the figure.
A SFR of $0.01\rm M_\odot\ yr^{-1}$ translates into
$L_{H\alpha}=2\times10^{39}\rm erg\ s^{-1}$ (assuming no extinction by dust),
which is a level of emission line luminosity that previous work reported.
Such star formation is likely spatially extended.
A careful removal of low-level star formation would be necessarily
to put a better constraint on the role of post-AGB stars.

To summarize, emission line luminosities due to post-AGB photo-ionization
seems to be a minor component of the overall luminosities.
However, other authors claim a substantial role of post-AGB stars
\citep{sarzi10,cidfernandes11,yan11b}.  This issue remains unresolved
and further effort is needed to settle it.
One difficulty that hampers a clear conclusion is our limited
understanding of the last stage of the stellar evolution.
Another difficulty is the lack of suitable observational data
to address the issue.  Dedicated observations would be essential.
For these difficulties, we cannot try to pursue the issue further
in this paper, but we should bear in mind the potentially non-negligible
contamination from post-AGB stars.
Given the unclear (but likely minor) role of 
post-AGB stars, we do not correct for their contamination to
the observed emission line luminosities throughout the paper.
We assume that the observed emission is entirely due to AGN and SF.

%------------------------------------------
\subsection{Comparisons with radio power}

%---------------------
\begin{figure*}
  \begin{center}
    \FigureFile(52mm,120mm){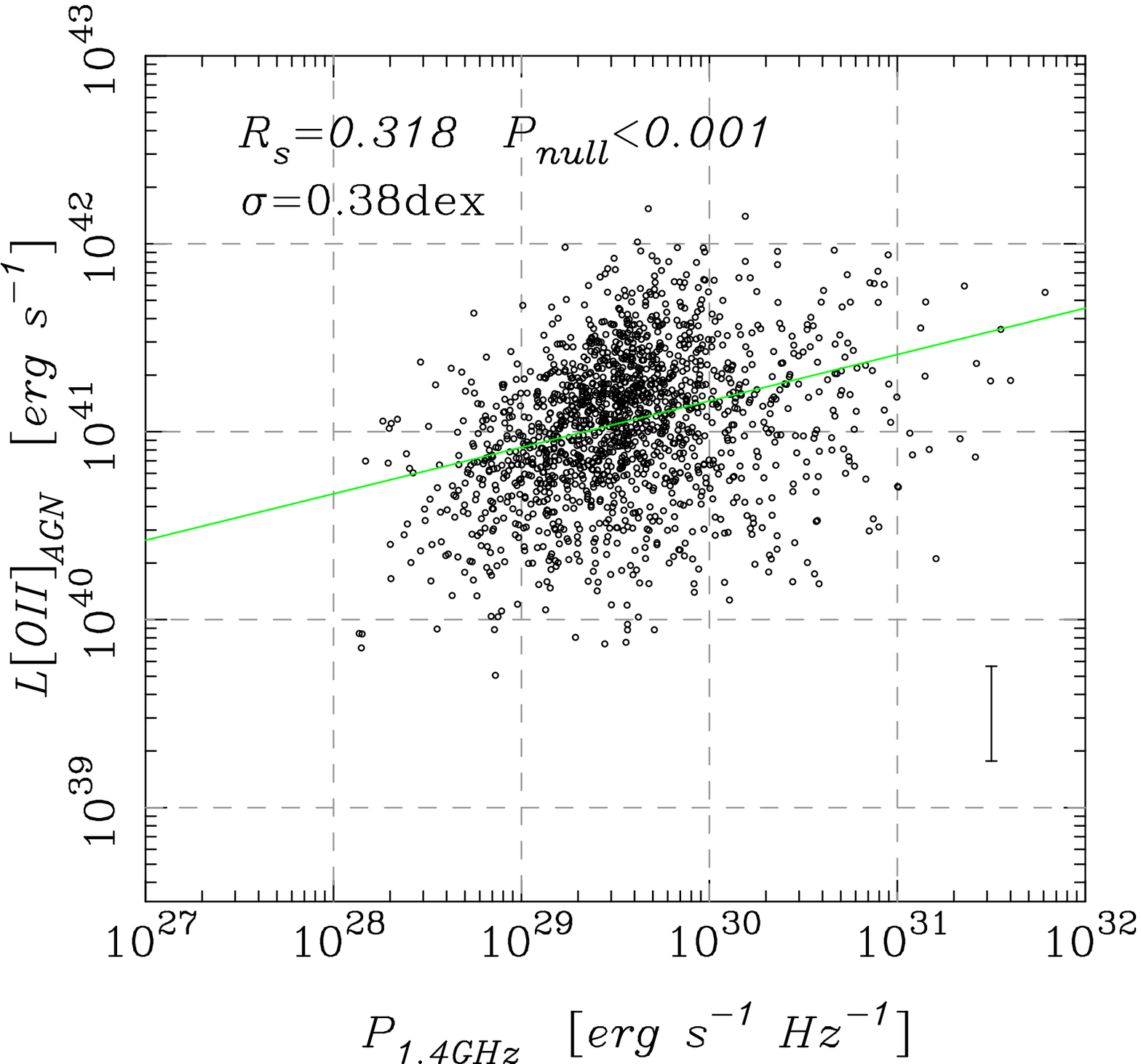}\hspace{0.5cm}
    \FigureFile(52mm,120mm){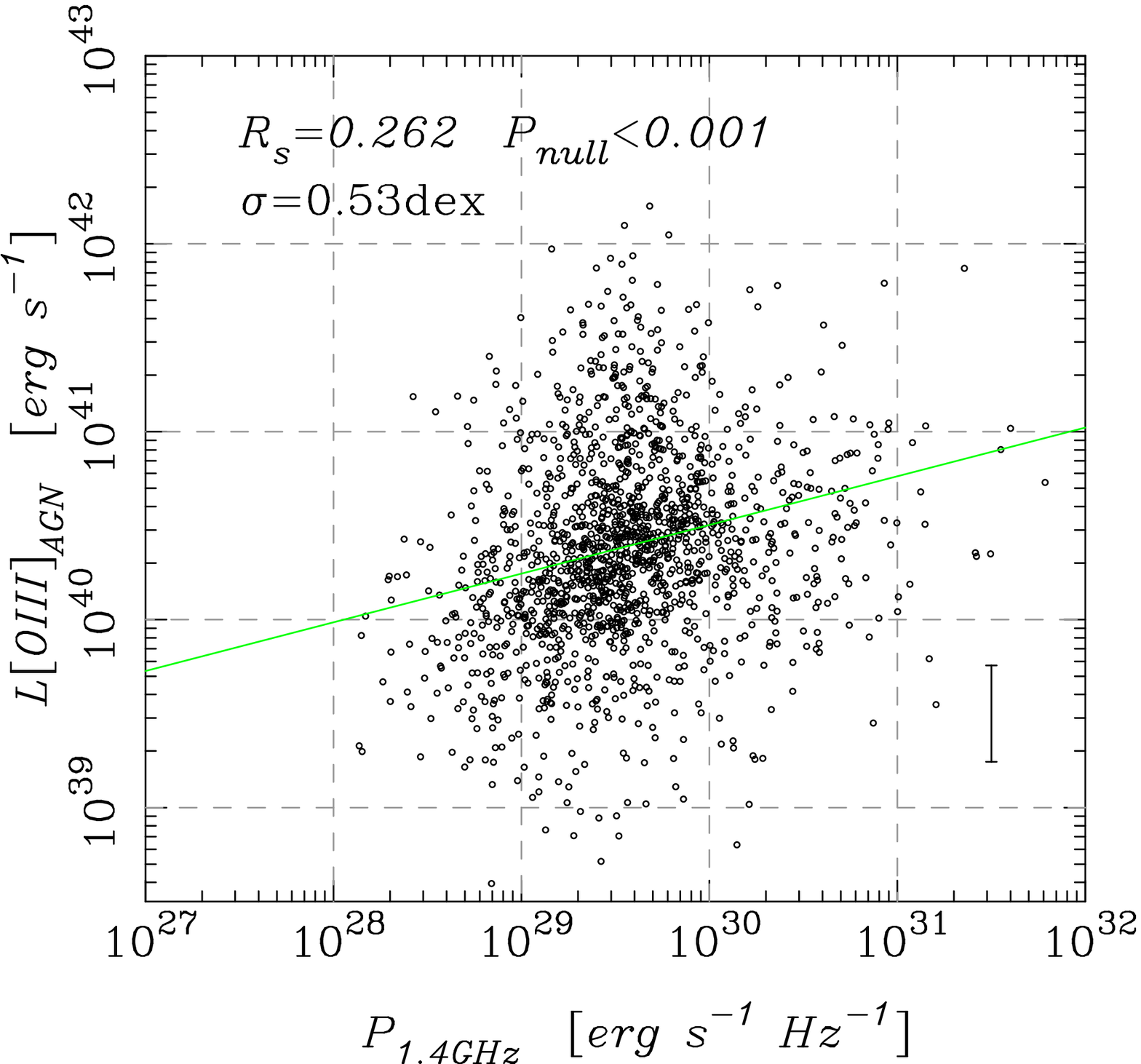}\hspace{0.5cm}
    \FigureFile(52mm,120mm){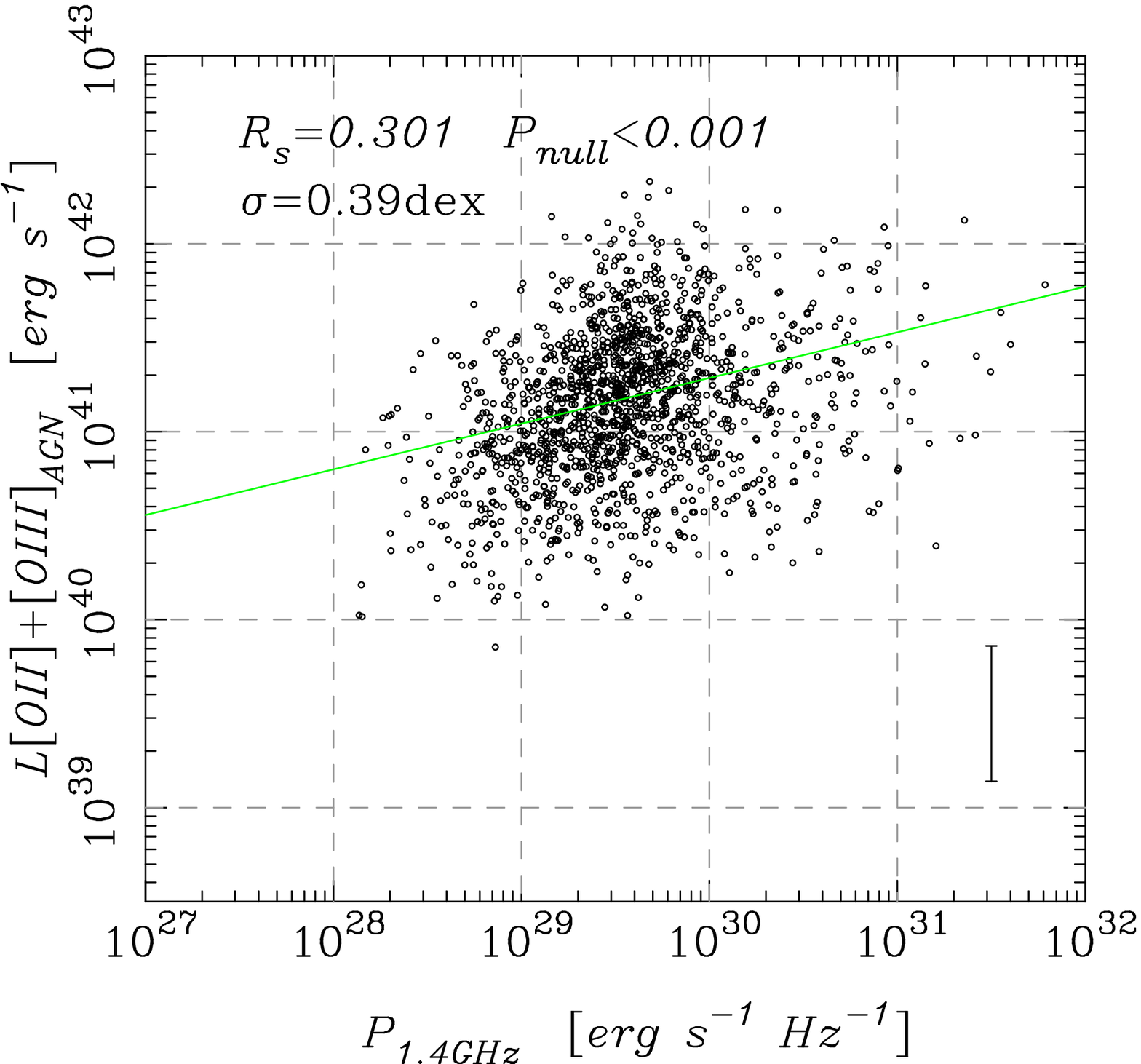}\\
  \end{center}
  \caption{
    As in Fig. \ref{fig:oiioiii_xray}, but here we compare
    optical emission line luminosities with radio power.  FIRST sources
    dominated by star formation are removed from the plot.
    The median error is shown on the bottom-right corner.
    The integrated FIRST fluxes are measured by fitting elliptical Gaussian
    to the sources, which makes it difficult to evaluate the flux
    uncertainties \citep{white97}.
    We do not quote the median uncertainty in the radio power.
  }
  \label{fig:oiioiii_radio}
\end{figure*}
%---------------------

We turn our attention to radio sources.
We cross-matched our objects with FIRST \citep{becker95,white97}
within 1 arcsec apertures
as described in Paper-I to compare optical power with radio power.
We miss extended radio lobes/jets, but
extended radio sources constitute only 10\% of the radio galaxies
and hence should not strongly affect our results here \citep{lin10}.
Also, extended radio lobes may not be a good measure of instantaneous
AGN activity.  We use $f_{integrated}$ from the FIRST catalog.
Our results remain the same if we use $f_{peak}$.

We plot  the optical power against radio power in Fig. \ref{fig:oiioiii_radio}.
Note that we remove radio sources that are dominated by
star formation from the comparisons (Fig. 13 of Paper-I shows
how we identify star formation dominated sources).
We observe a weak correlation between optical and radio
and the Spearman's rank test rejects the null hypothesis.
The weak correlation is in contrast to X-rays, which show stronger correlations.
The loose radio-optical correlation may appear inconsistent with
previous studies (e.g., \cite{saunders89,baum89,rawlings91,hes93,ho99}).
These authors observed a relatively tight correlation between optical emission
and radio power over a range of AGN types.

We do not consider our finding inconsistent with earlier findings for
two reasons.  First, our radio galaxy sample has more than an order
of magnitude lower radio power than most of the previous studies.
Majority of the Oxygen-excess objects are low-luminosity AGNs, while
the previous studies focused on powerful radio galaxies.
We are therefore not looking at the same population.
Second, we exclude sources that are strongly contaminated
by underlying star formation activities, while the previous studies did not.
We find that the correlations slightly tighten up if we include those sources because
we include the strong SFR-radio power correlation in the statistics.
Our result may still not be consistent with observations by \citet{ho99},
who studied nearby low-luminosity AGNs in early-type galaxies, in which
the underlying star formation is probably not very strong.
However, their early-type galaxy sample shows a large scatter comparable
to ours (see their Fig. 2).  
For these reasons, we do not consider that our results are in strong
conflict with previous studies.
Recently, \citet{degasperin11} observed no clear
correlation between radio and X-ray powers.  This probably suggests 
that there is no strong correlations between radio and optical powers
in consistent with our result here.

The positive correlation shows the black hole activities and radio
activities are weakly correlated in the Oxygen-excess objects under study.  
We note that 
several objects are detected both in X-rays and radio.
We compare the X-ray and radio powers and find no significant correlations.
Our statistics is very poor, but the observed lack of correlation
is consistent with \citet{degasperin11}.
The optical and X-ray powers are likely a good indicator of on-going AGN
activities, while the radio power is likely not.

%------------------------------------------
\subsection{Is {\sc [oii]} an indicator of SFRs of AGN host galaxies?}

%---------------------
\begin{figure*}
  \begin{center}
    \FigureFile(53mm,80mm){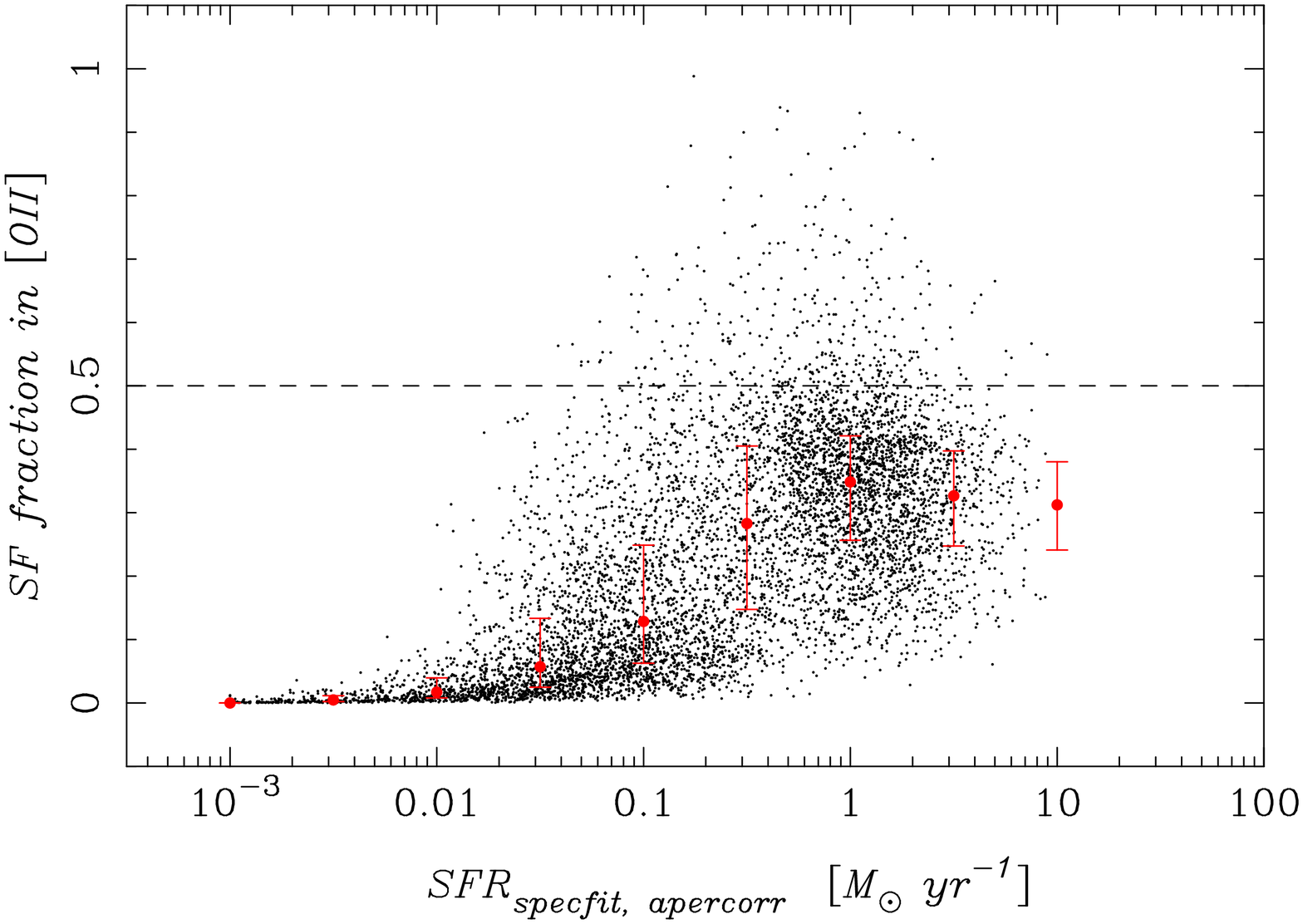}\hspace{0.2cm}
    \FigureFile(53mm,80mm){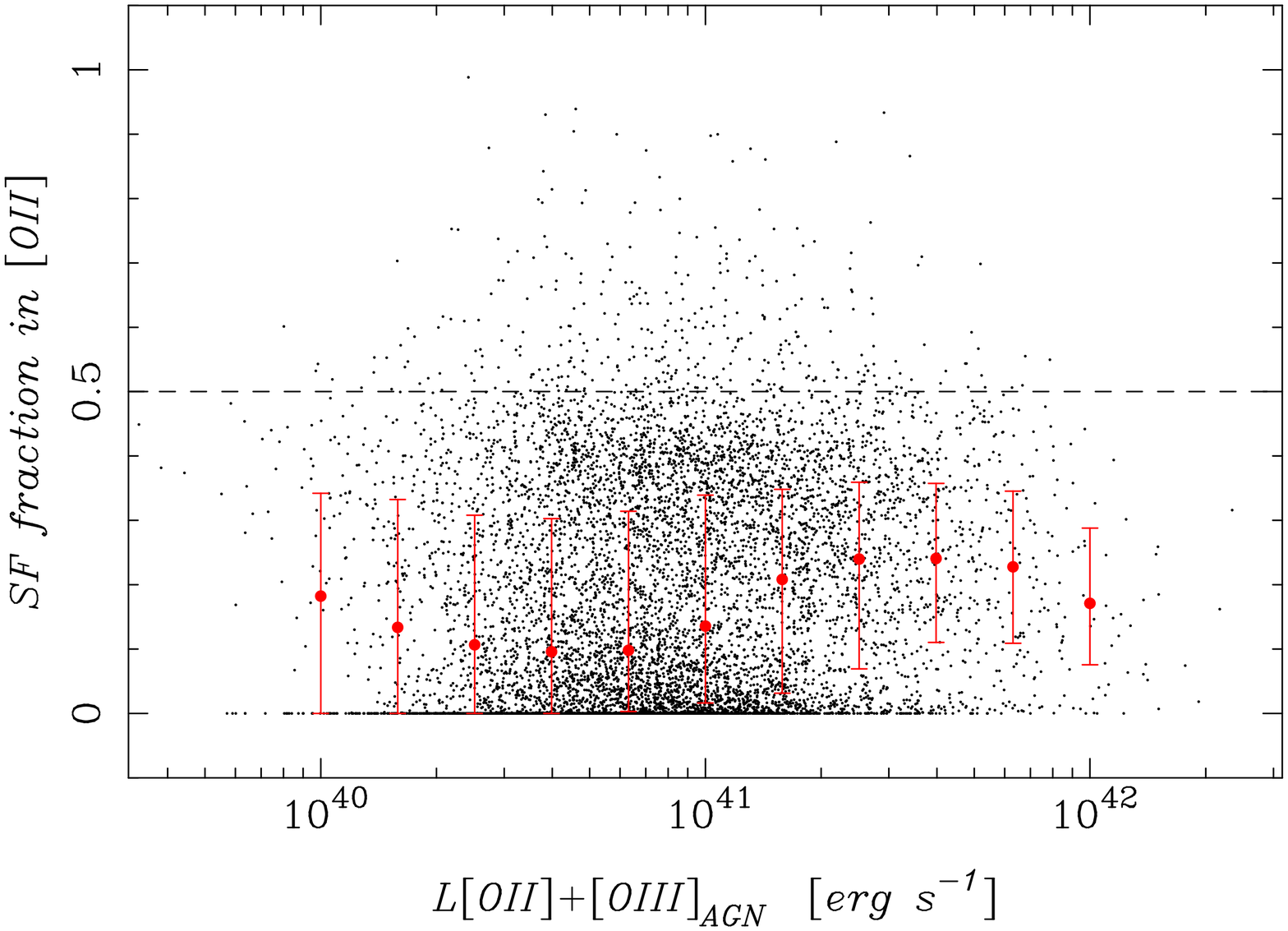}\hspace{0.2cm}
    \FigureFile(53mm,80mm){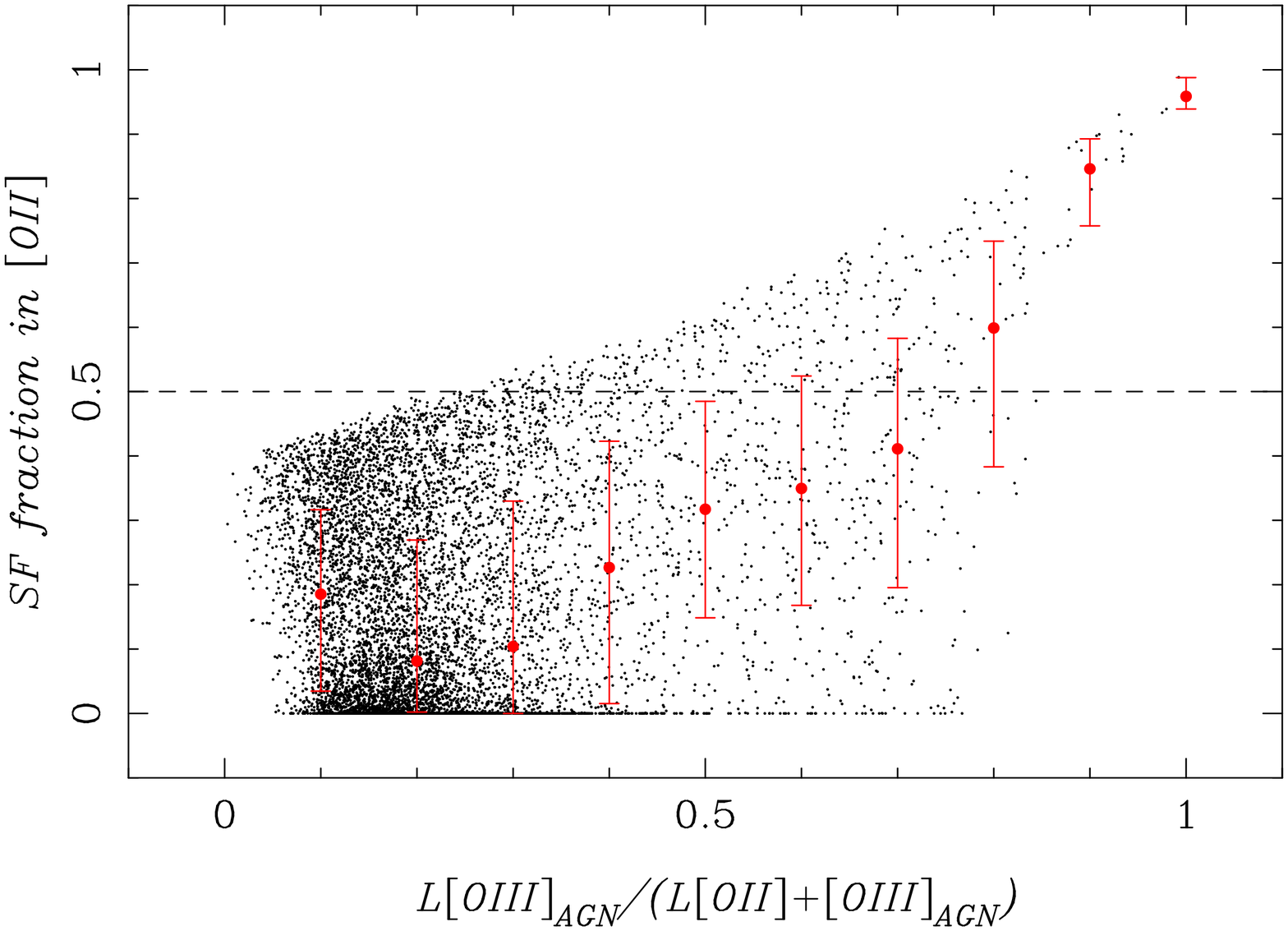}
  \end{center}
  \caption{
    {\it Left:} Fractional contribution of {\sc [oii]} due to star formation to the overall {\sc [oii]}
    luminosity plotted against SFR.  The dashed line shows the equality of SFR and AGN
    components.  Star formation dominates above the line and AGN dominates below the line.
    The points with error bars show the median and quartiles of the distribution.
    {\it Middle:} Same as the left panel, but here the horizontal axis is AGN power.
    {\it Right:} Same as the left panel, but the horizontal axis is ionization state.
    {\sc [oii]}$<${\sc [oiii]} holds at {\sc [oiii]/([oii]+[oiii])}$>0.5$,
    and {\sc [oii]}$>${\sc [oiii]} holds at {\sc [oiii]/([oii]+[oiii])}$<0.5$.
  }
  \label{fig:oii_frac}
\end{figure*}
%---------------------

The {\sc [oii]} emission is sometimes used as a proxy for host galaxy SFRs of 
high ionization AGNs such as Seyferts because this line is not very strong in those AGNs
(e.g., \cite{ho05}).
It may be reasonable to assume that {\sc [oii]} is due to star formation in Seyfert galaxies,
but is it still good for LINERs?
The newly developed method allows us to statistically separate
star formation and AGN components of a given emission line.
We take this opportunity to quantify how well {\sc [oii]} is
correlated with SFRs over a range of AGN power and ionization states.

Fig. \ref{fig:oii_frac} shows the fractional contribution of {\sc [oii]}
due to star formation to the overall {\sc [oii]} luminosity as a function of
SFRs of the host galaxies (left panel), AGN power (middle panel),
and ionization level (right panel).
We do not have a lot of Oxygen-excess galaxies dominated by
star formation (i.e., those above the dashed line), but this is
simply a selection bias that we cannot find weak AGNs in
actively star forming galaxies because our method uses a contrast
between $L_{SF}$ and $L_{AGN}$.  If {\sc [oii]} was a perfect indicator
of star formation, we would expect that galaxies line up at
$L_{[OII],SF}/(L_{[OII],SF}+L_{[OII],AGN})=1$.

Starting with the left panel, {\sc [oii]} due to AGN completely
dominates in galaxies with low SFRs, while the star formation component
comes in at moderate-high SFRs.  This is a totally expected behavior.
The observed {\sc [oii]} does not give any sensible SFRs in quiescent galaxies.
Even in actively star forming galaxies ($SFR_{specfit}\sim1\rm \ M_\odot\ yr^{-1}$),
{\sc [oii]} due to star formation does not dominate, although
we are limited by the selection bias there.

One might expect that the AGN component becomes negligible
in weak AGNs and {\sc [oii]} might be a good indicator of SFRs.
But, the middle plot in Fig. \ref{fig:oii_frac} shows that
this is not the case.  As we will discuss in the next section,
there is a positive correlation between the AGN power and underlying
star formation activities.  Due to this correlation, there is
no strong dependence between the dominance of SF/AGN component
and AGN power.
There is always a fair chance of suffering from AGN contribution
in the observed {\sc [oii]} luminosity.

Finally, the right panel shows the dominance of the star formation component
as a function of ionization level.  Most of the LINERs ({\sc [oii]$_{AGN}$}$>${\sc [oiii]$_{AGN}$})
are dominated by AGN emission and {\sc [oii]} due to star formation dominates
only in the very high ionization galaxies.  For the vast majority of low-luminosity AGNs,
{\sc [oii]} is unlikely a good indicator of star formation.

%---------------------
\begin{figure*}
  \begin{center}
    \FigureFile(70mm,80mm){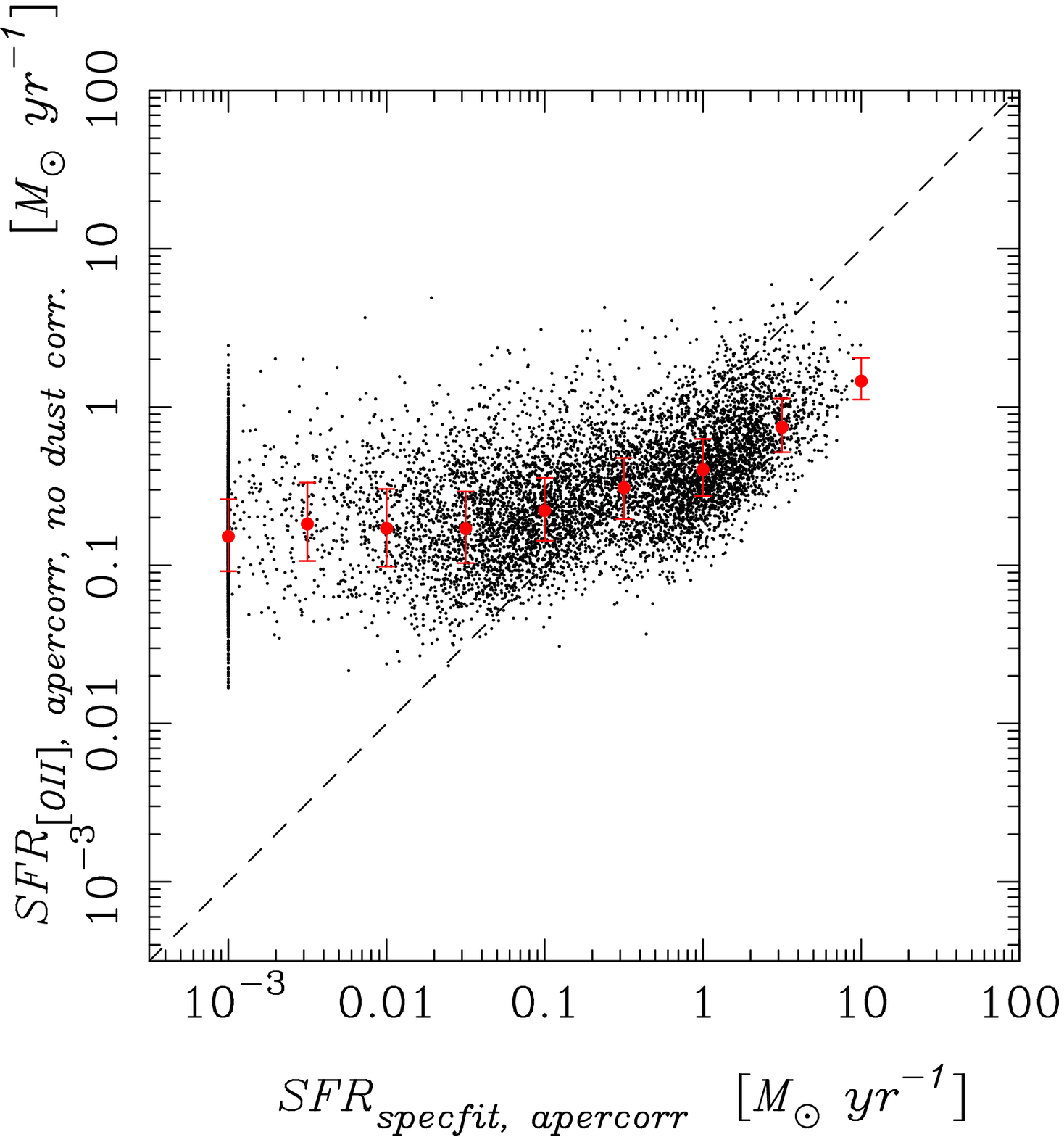}\hspace{0.5cm}
    \FigureFile(70mm,80mm){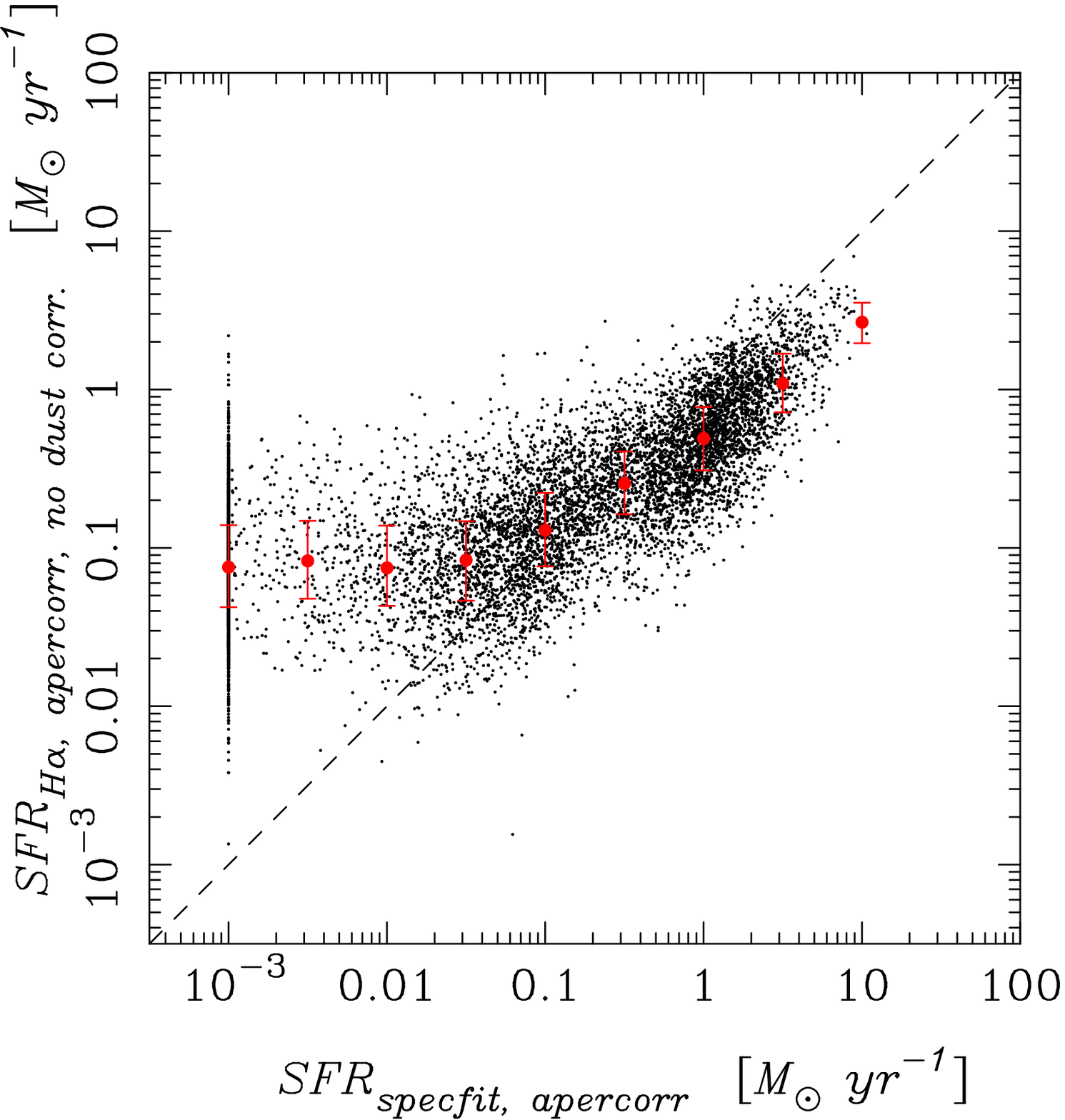}\\
  \end{center}
  \caption{
    SFRs from {\sc [oii]} (left) and H$\alpha$ (right) plotted against
    SFRs from the spectral fits.
    We only correct for the fiber aperture loss and do not apply
    any further correction to {\sc [oii]} and H$\alpha$
    (i.e., no subtraction of AGN component and no correction for dust)
    and simply convert the observed luminosity into SFRs.
    The points with error bars show the median and quartiles of the distribution.
  }
  \label{fig:oii_frac5}
\end{figure*}
%---------------------

To be further quantitative, let us ask a simple question ---
if we assume that {\sc [oii]} is entirely due to star formation,
how well can we reproduce the SFRs of the hosts?
Left panel in Fig. \ref{fig:oii_frac5} gives the answer.
We simply translate an observed {\sc [oii]} luminosity into SFR using the formula
given in \citet{kennicutt98} without any subtraction of the AGN component or
any dust correction.
As can be seen, the SFRs from {\sc [oii]} do not correlate
with true SFRs very well.
In particular, one grossly overestimates SFRs in quiescent galaxies.
In those galaxies, star formation is very weak and the observed {\sc [oii]}
is nearly entirely due to AGN (Fig. \ref{fig:oii_frac} left panel).
The assumption of {\sc [oii]} 
due to star formation completely fails here.
Another point to mention is that the systematic offset seen at 
$\sim1\rm M_\odot\ yr^{-1}$ is mostly due to the ignorance of dust.
If we correct for $\tau_V=1$ extinction, which is typical 
in star forming regions \citep{hopkins03}, the systematic offset
is largely gone.  However, the dust correction increases
the discrepancy in low SFR galaxies because it shift all the
points upwards.

As shown in Paper-I, the collisionally excited {\sc [oii]} and {\sc [oiii]}
suffer (or benefit when we identify AGNs using these lines) from 
AGN contribution, while H$\alpha$ is less affected.  This is consistent
with what the BPT diagram tells us --- AGNs show stronger collisionally
excited lines than the Balmer lines and that is why galaxies form
a distinct sequence from the star formation sequence to the top-right
corner of the diagram, not to the bottom-left corner.
Due to the relatively weak AGN contribution to H$\alpha$,
SFRs from H$\alpha$ indeed better agree with those
from the spectral fits as shown in the right panel of Fig. \ref{fig:oii_frac5}.
Therefore,
H$\alpha$ is a better indicator of SFRs in AGNs than {\sc [oii]}.
However, H$\alpha$ still largely overestimates SFRs in quiescent galaxies
by more than an order of magnitude.

To sum up, one should be cautious about using {\sc [oii]}
as an indicator of SFRs in AGNs.  It can give SFRs that
are offset by more than an order of magnitude.
In particular, we suggest that one should not use {\sc [oii]}
when a host galaxy has red colors.  The galaxy may be a dusty galaxy,
but it may well be passive galaxies, for which {\sc [oii]}
is entirely due to AGN.
As a result, one wildly overestimates SFRs.
Also, dust extinction needs to be accounted for.
%, although
%constant extinction correction would result in grossly overestimated
%SFRs in quiescent galaxies.
One should use continuum information whenever available because
the spectral fitting is likely to deliver better SFRs due to
less AGN contamination to the continuum spectrum \citep{schmitt99}.
It may often be the case when the continuum spectrum
is not available with sufficient signal-to-noise ratios
especially at high redshifts.
Recent availability of multi-wavelength photometry would give
a way around it.
SED fits of multi-wavelength broad-band photometry 
seem to give reasonable SFR estimates (e.g., \cite{santini09}).
If one is careful enough to remove photometry that can be affected
by AGN component (UV continuum or IR thermal emission) from the SED fits,
it would provide better SFR estimates than {\sc [oii]} and H$\alpha$.
We shall emphasize that the key is to use stellar light only.
It would be challenging to estimate SFRs of QSOs,
whose continuum is dominated by AGN emission, not by stars.
If QSO continuum can be subtracted by assuming a power-law form,
one can fit the spectrum of the remaining stellar component.
An accuracy of SFRs from such a fit remains to be quantified.

%------------------------------------------
\section{Nature of the Oxygen-excess objects: AGN activities and the host galaxy properties}

Following the development of the basis to characterize AGN activities,
we now move on to discuss the nature of Oxygen-excess objects
with an emphasis on the relationship between the AGN and host galaxy properties.
We start with the fraction of AGNs as a function of host galaxy properties,
followed by a detailed look at a partition between LINERs and Seyferts.
We then show scatter plots of the host galaxy properties and black hole masses.
Finally, we study the AGN activity and compare the black hole growth rate
and the host galaxy growth rate.

%------------------------------------------
\subsection{Dependence of AGN fraction on Stellar mass, Color, SFR and Morphology}

%---------------------
\begin{figure}
  \begin{center}
    \FigureFile(80mm,80mm){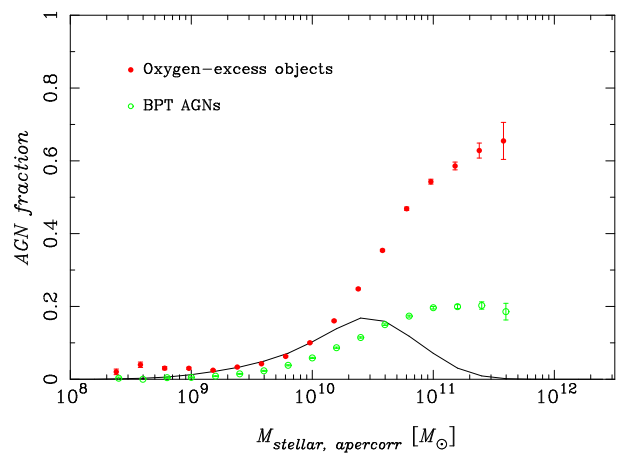}
  \end{center}
  \caption{
    Fraction of AGNs plotted against stellar mass.
    The filled and open circles are Oxygen-excess objects and AGNs from the BPT
    diagram, respectively.
    The solid line shows the fractional distribution of all the galaxies
    in our sample.
  }
  \label{fig:agn_comp1}
\end{figure}
%---------------------

Essentially all galaxy properties depend on mass of galaxies and we
start with stellar mass dependence of the AGN fraction 
in Fig. \ref{fig:agn_comp1}.
The AGN fraction is a very strong function of stellar mass of
host galaxies.  Most AGNs that we have identified reside in galaxies
with $M_{stellar, apercorr}>10^{10}\rm M_\odot$.  In the most massive galaxies
with $M_{stellar, apercorr}>10^{11}\rm M_\odot$, 60\% of the galaxies harbor AGNs.
We note that this is just a lower limit.  A large fraction
of galaxies in our sample do not exhibit strong enough Oxygen lines
to apply our method (i.e., OnBn defined in Paper-I) and the stacked spectrum
of those galaxies exhibits LINER-like line ratios.  Therefore, many of
the unclassified galaxies may well be AGNs (see also discussions
on the non-AGN photo-ionization in Paper-I and earlier sections of this paper).  
In contrast to the Oxygen-excess method,
the BPT diagnostics gives an AGN fraction of only 20\% in massive galaxies.
This demonstrates the sensitivity of our method to identify AGNs.

\citet{heckman04} found that the fraction of active black holes (i.e., AGNs)
is highest in intermediate mass black holes and it 
decreases in the most massive black holes based on BPT AGNs.
We have replaced the stellar mass with black hole mass (see below for
the derivation of the black hole mass) in Fig. \ref{fig:agn_comp1} and
found that a fraction of Oxygen-excess objects continues to increase towards
the high mass end.
This difference from \citet{heckman04} is due to the difference in the redshift ranges explored.
\citet{heckman04} did not apply any redshift selection to the main galaxy
sample \citep{strauss02}.  Due to the flux limited nature of the sample,
an average redshift of galaxies is higher for more massive galaxies and
therefore an average sensitivity to emission line luminosity is lower.
This resulted in a reduced sensitivity to identify AGNs in very massive galaxies and
caused a decline in the fraction of active black holes at the high mass end.
We have confirmed that we reproduce the \citet{heckman04}'s trend by
changing the redshift range to $0<z<0.3$ and redoing the whole analysis.

%---------------------
\begin{figure*}
  \begin{center}
    \FigureFile(80mm,80mm){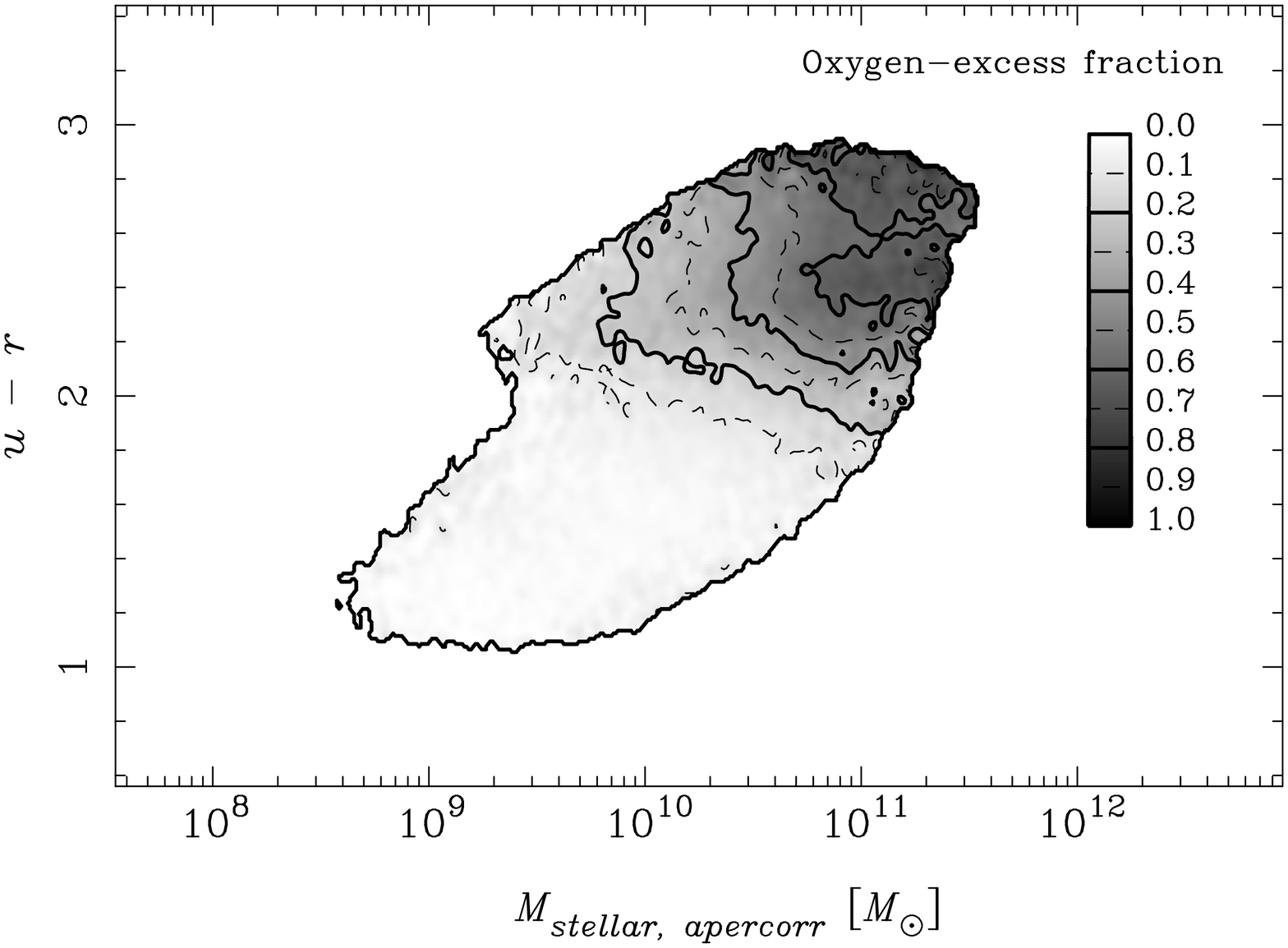}\hspace{0.5cm}
    \FigureFile(80mm,80mm){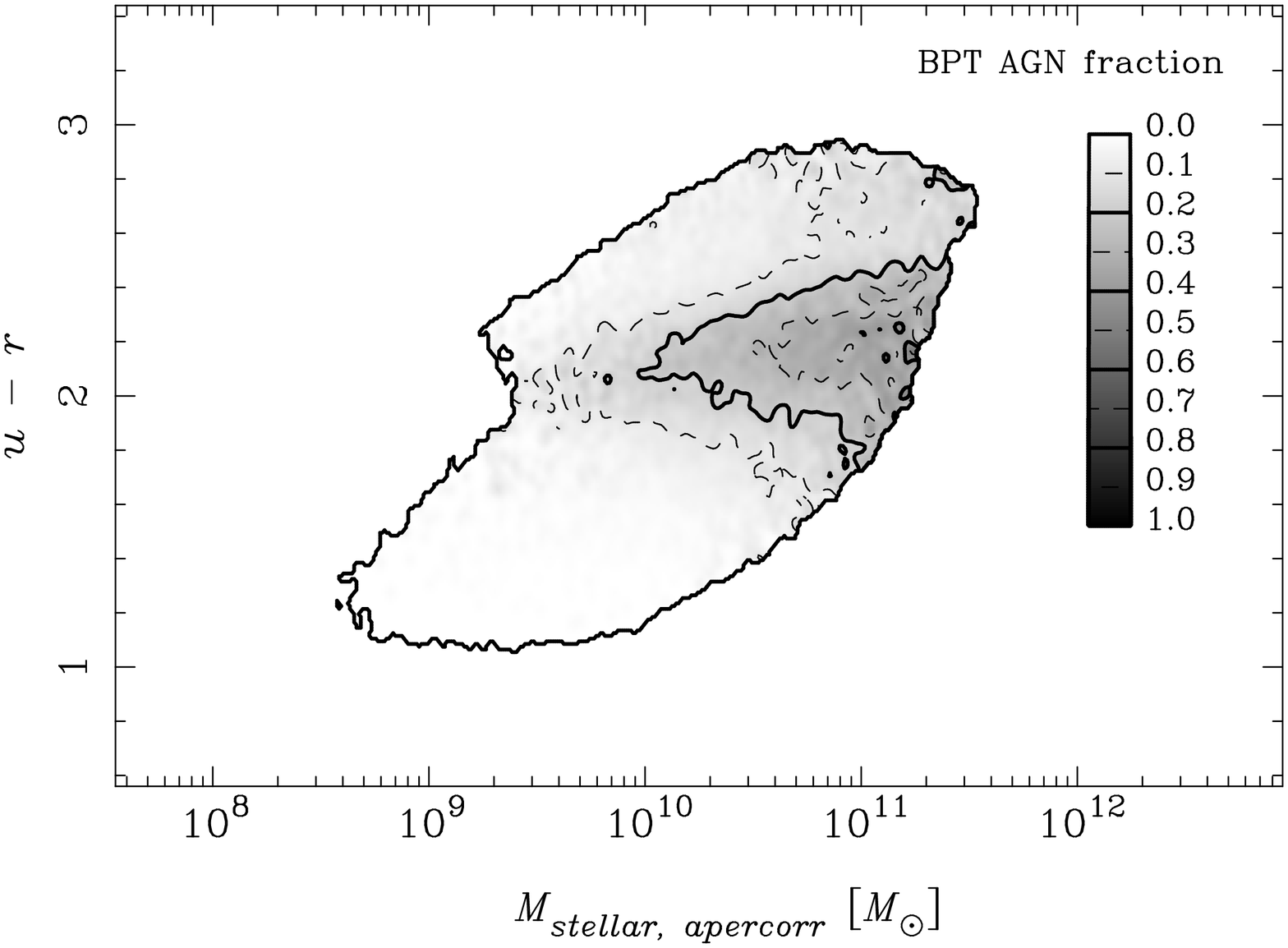}
  \end{center}
  \caption{
    {\bf Left:}
    Fraction of Oxygen-excess objects as functions of rest-frame $u-r$ color and stellar mass.
    The solid contours show the fractions of 0, 20, 40, 60, 80, and 100\%,
    and the dashed contours show 10, 30, 50, 70, and 90\%.
    {\bf Right:}
    As in the left panel, but here we show the fraction of BPT AGNs.
  }
  \label{fig:agn_comp_color}
\end{figure*}
%---------------------

We further discuss color, SFR and morphology dependence of the AGN fractions.\\

\noindent
{\bf Color (Fig. \ref{fig:agn_comp_color}):}
We compute rest-frame $u-r$ color from the model magnitudes \citep{stoughton02}
using the $k$-correction code by \citet{blanton07} and plot
the AGN fraction as functions of stellar mass and rest-frame $u-r$ color.
We find that Oxygen-excess objects prefer red galaxies ($u-r\gtrsim2$) at all masses.
As expected from the stellar mass dependence in Fig. \ref{fig:agn_comp1},
the fraction increases from low to high mass galaxies.
The most massive galaxies show the highest AGN fraction, reaching to nearly 70\%.
The fraction decreases towards bluer colors, but this is at least
partly driven by the selection bias that we cannot identify weak AGNs
in actively star forming galaxies.

In contrast to the clear preference of Oxygen-excess objects to red colors,
a large fraction of BPT AGNs reside in the green valley around $u-r\sim2$
in consistent with \citet{kauffmann03} and \citet{schawinski10}.
This is due to the fact that BPT is not very sensitive to
AGNs in passive galaxies.  As shown later, most AGNs in passive galaxies are
low-luminosity AGNs.  They often show too weak H$\beta$ to
apply the BPT method (see the stacked spectrum of O+Bn in Fig. 7 of Paper-I).
As a result, the BPT misses a significant fraction of low-luminosity AGNs
on the red sequence, and that causes the large difference in the AGN
fraction in red galaxies between Oxygen-excess and BPT.
In other words, BPT AGNs in the green valley are relatively strong AGNs.
We note that objects detected in radio or X-ray do not show
a clear excess in the green valley.
\\

%---------------------
\begin{figure*}
  \begin{center}
    \FigureFile(80mm,80mm){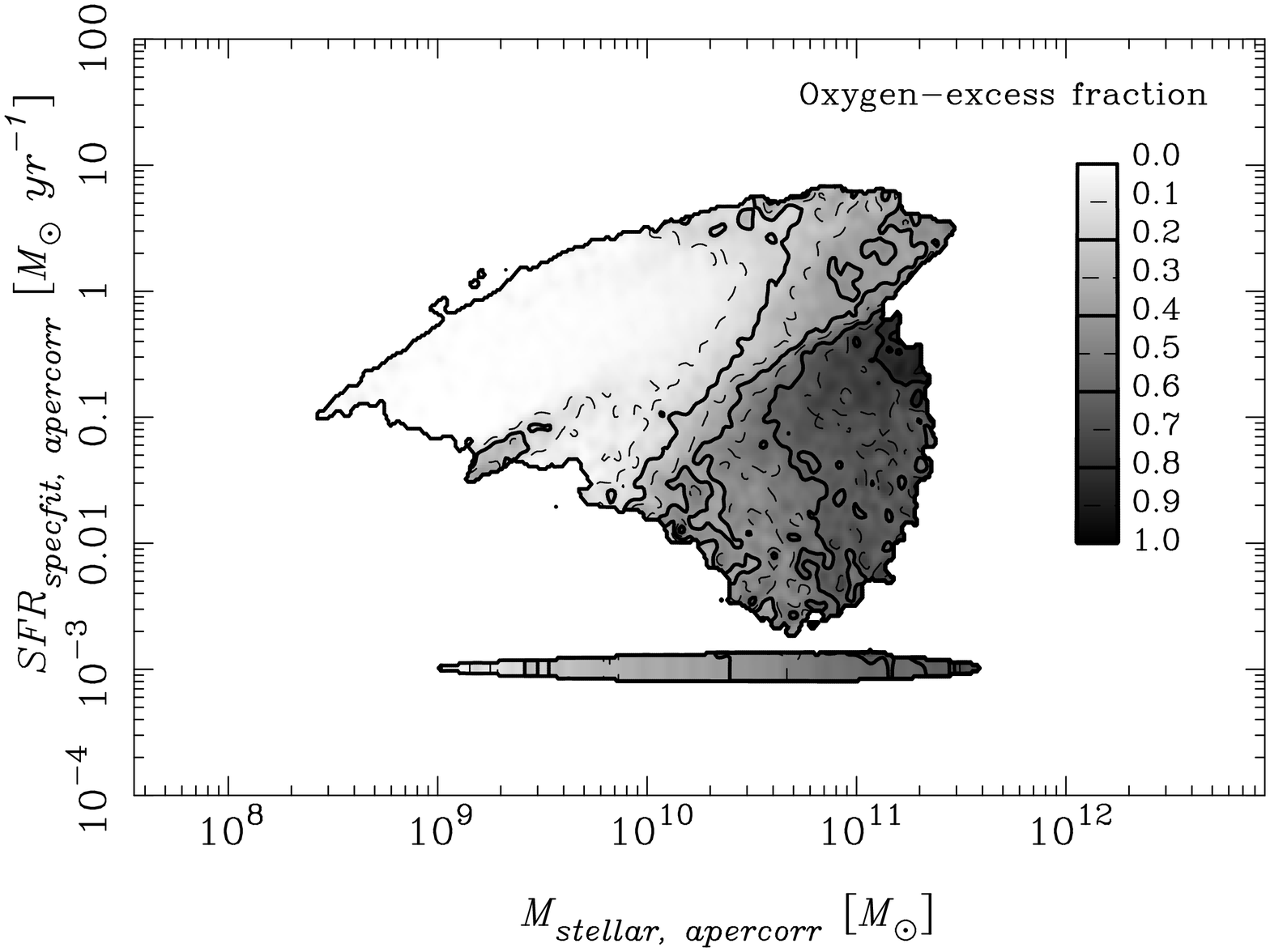}\hspace{0.5cm}
    \FigureFile(80mm,80mm){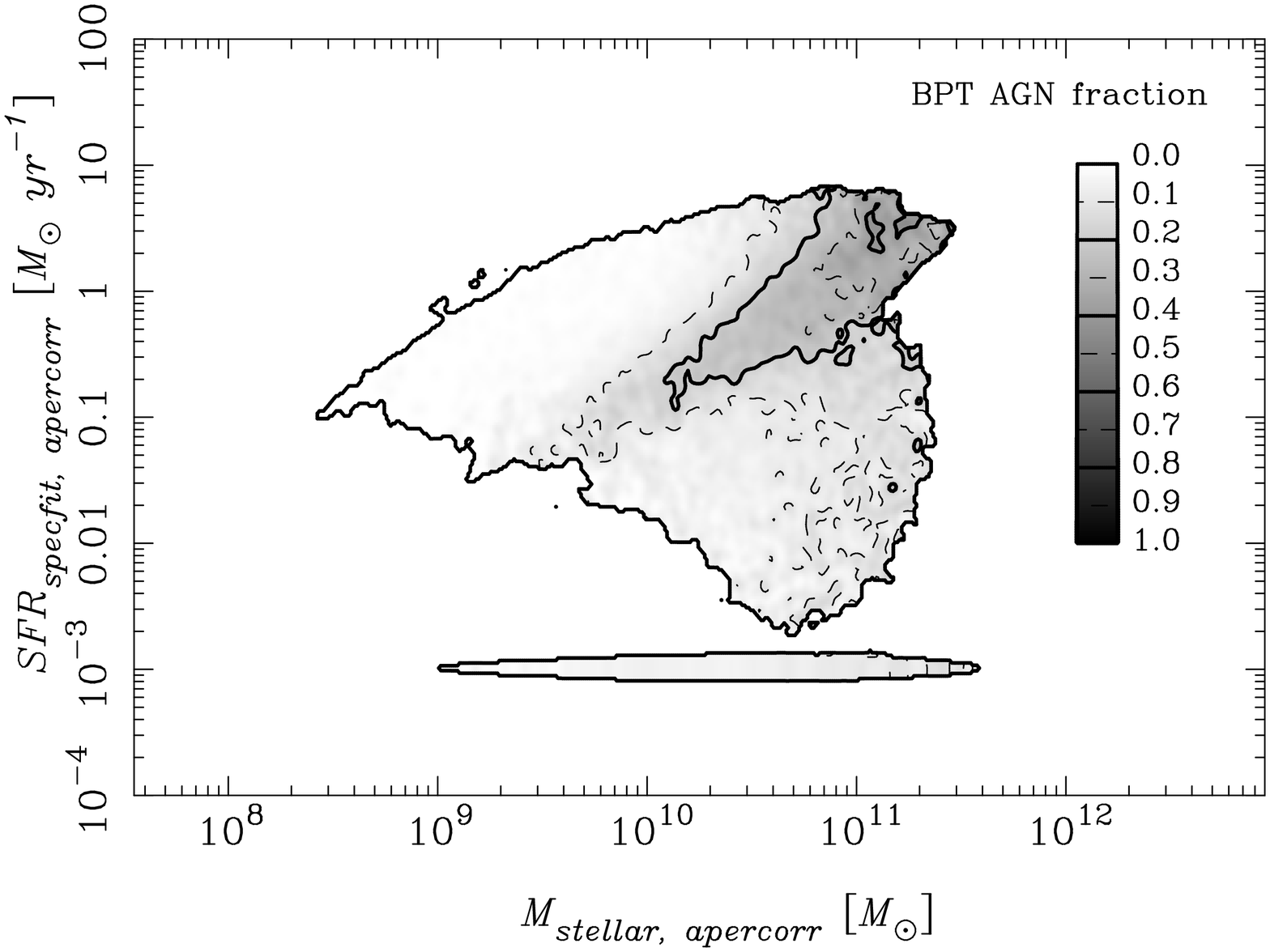}
  \end{center}
  \caption{
    {\bf Left:}
    Fraction of Oxygen-excess objects as functions of SFR and stellar mass.
    The contours are the same as in Fig. \ref{fig:agn_comp_color}.
    Galaxies with SFRs$<10^{-3}\rm M_\odot\ yr^{-1}$ are summed up at
    SFR$=10^{-3} \rm M_\odot\ yr^{-1}$.
    {\bf Right:}
    As in the left panel, but for BPT AGNs.
  }
  \label{fig:agn_comp_sfr}
\end{figure*}
%---------------------

\noindent
{\bf SFR (Fig. \ref{fig:agn_comp_sfr}):}
In addition to the color, SFR is also one of the most important properties of galaxies.
We quantify the AGN fraction as a function of SFRs
and stellar mass of the host galaxies.  Low-mass galaxies are predominantly
star forming, but massive galaxies exhibit bimodal distribution
of SFRs.
The BPT AGN fraction shown in the right panel is constantly low at
$\rm SFR\lesssim0.1 M_\odot\ yr^{-1}$, but it shows a rapid increase at higher SFRs.
On the other hand, the Oxygen-excess objects show 
a contrasting behavior.  We interpret it by a combination
of two effects.  At low SFRs, the Oxygen-excess method is
fairly sensitive and it gives a much higher AGN fraction than BPT.
Because the method is based on identifying excess Oxygen
luminosity, it becomes less efficient as underlying
star formation becomes more active.  This causes the drop of
the AGN fraction around  $1\rm M_\odot\ yr^{-1}$, where significant star
formation component kicks in (see Fig. \ref{fig:oii_frac}).
This plot shows the sensitivity function of the Oxygen-excess method
(we will further quantify it below).
Its sensitivity is similar to BPT when SFRs are high, but it is much more
sensitive at low SFRs.  The Oxygen-excess method is suited
to study AGNs in quiescent galaxies.

The fraction of Oxygen-excess objects seems to more strongly
depend on stellar mass than on SFR.  The fraction of Oxygen-excess objects
with $\rm SFR\sim0.1\ M_\odot\ yr^{-1}$ drops from 70\% at high mass
to nearly 0\% at low mass.
Mass is likely the primary parameter to trigger the AGN activity.\\

%---------------------
\begin{figure*}
  \begin{center}
    \FigureFile(80mm,80mm){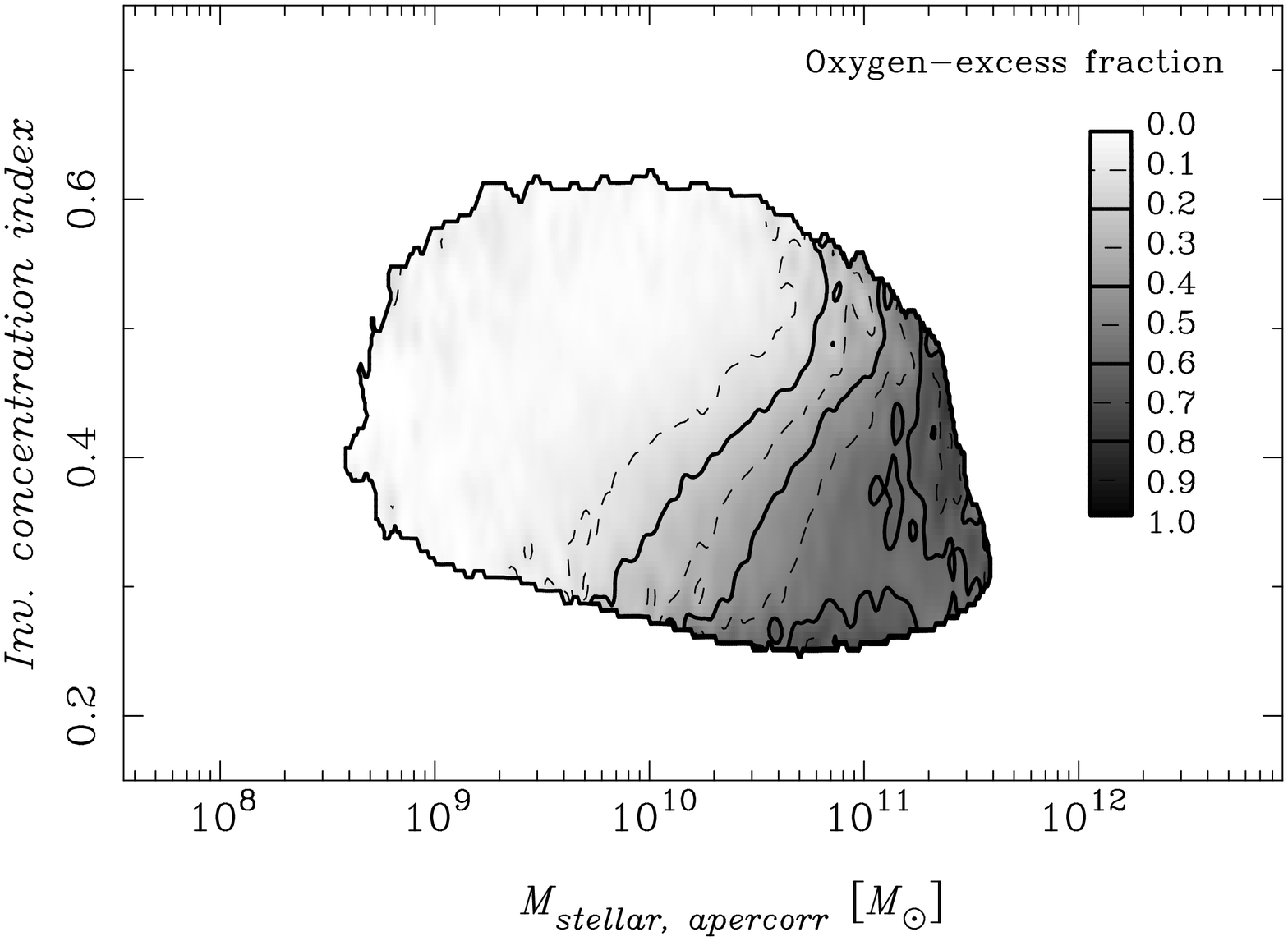}\hspace{0.5cm}
    \FigureFile(80mm,80mm){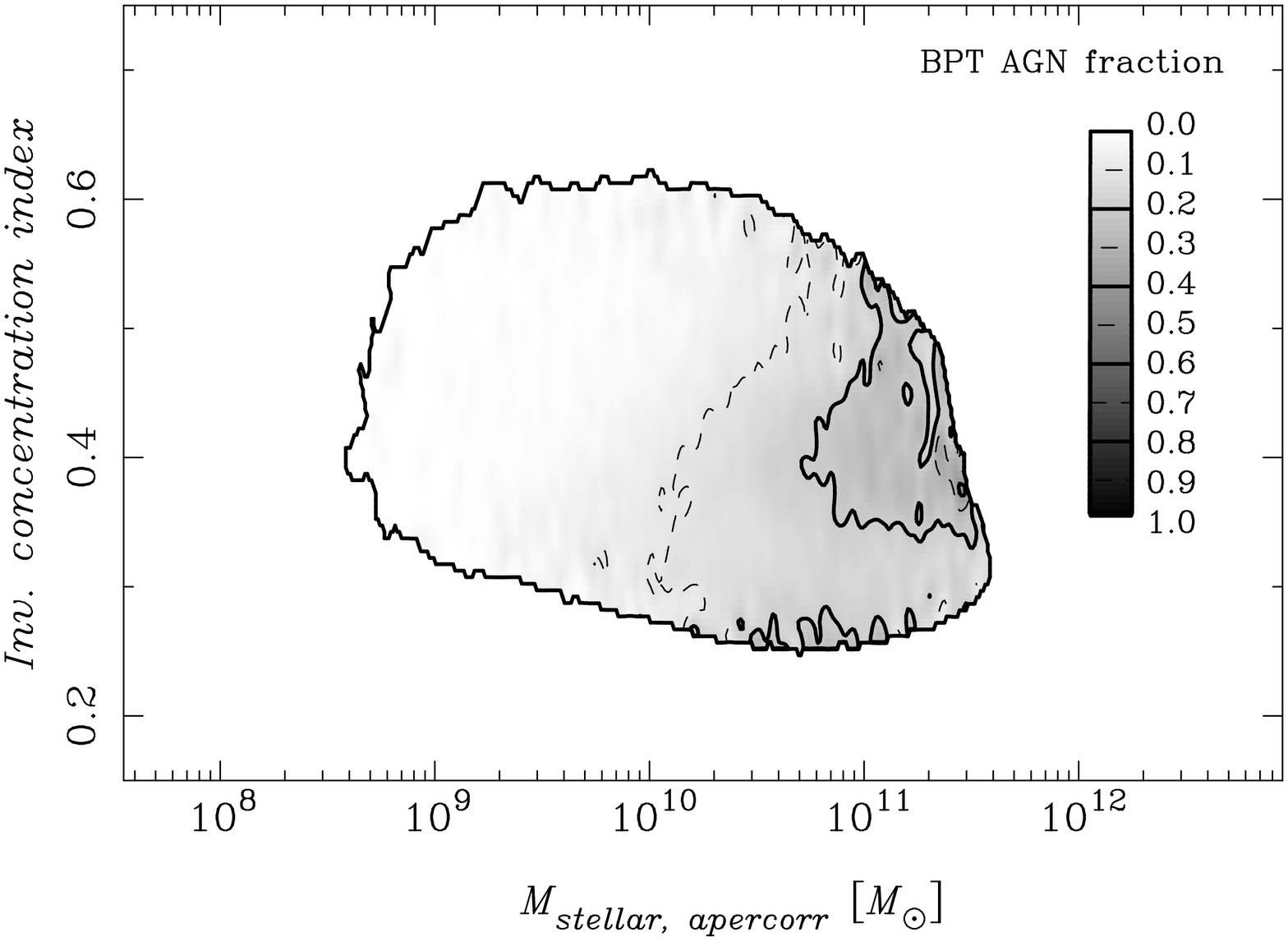}
  \end{center}
  \caption{
    {\bf Left:}
    Fraction of Oxygen-excess objects as functions of inverse concentration index
    measured in the $z$-band and stellar mass.
    The contours are the same as in Fig. \ref{fig:agn_comp_color}.
    {\bf Right:}
    As in the left panel, but for BPT AGNs.
  }
  \label{fig:agn_comp_cin}
\end{figure*}
%---------------------

\noindent
{\bf Morphology (Fig. \ref{fig:agn_comp_cin}):}
As an indicator of morphology, we use the inverse concentration index 
\citep{shimasaku01,strateva01} defined by $R_{50}/R_{90}$, where
$R_{50}$ and $R_{90}$ are the radii which enclose 50\% and 90\% of
the total light (Petrosian flux; \cite{petrosian76,stoughton02}) in the $z$-band
to minimize effects of on-going star formation.
 The inverse concentration index is of course
affected by the seeing.  The median seeing in the $z$-band is 1.3 arcsec.
We have confirmed that the trend in the plot does not change whether we
use only objects observed under good seeing or not.

We quantify the AGN fraction as functions of the inverse concentration index
and stellar mass in Fig.  \ref{fig:agn_comp_cin}.
Oxygen-excess objects tend to be massive early-type galaxies as expected
from the color-stellar mass dependence discussed above.
The Oxygen-excess fraction shows a mild dependence on morphology, but
again, stellar mass seems to be the primary parameter to trigger
AGN activity.  It is interesting to note that even if galaxies exhibit
early-type morphology, they do not show a hint of AGN activity at low mass.
In contrast to Oxygen-excess objects,  BPT AGNs prefer intermediate
morphology as they are strongly biased towards green valley galaxies
due to the poorly sampled low-luminosity AGNs.\\

Overall, Oxygen-excess objects tend to be massive red galaxies.
The fraction of Oxygen-excess objects decreases towards blue, star-forming,
late-type galaxies at a given stellar mass, but this is at least partly
driven by the selection bias that we miss weak AGNs in actively star forming galaxies.
On the other hand, the distribution of BPT AGNs is strongly skewed towards
galaxies in the green valley.  This is driven by the bias that
BPT diagnostics is not very sensitive to low-luminosity AGNs on the red sequence.
The figures presented in this section seem to suggest that BPT diagnostics does not
provide an unbiased sample of AGNs.  BPT AGNs are strongly biased.
%This bias should carefully be accounted for when BPT AGNs are used in
%any sort of galaxy studies.
  Oxygen-excess objects are also biased, but
the selection bias is relatively straightforward to quantify as shown below because
it is simply based
on a contrast between AGN power and underlying star formation
(it can be complicated when the featureless continuum emission from
AGN significantly contributes to the overall spectra).
We will carefully account for such biases in the following discussions.

%------------------------------------------
\subsection{Dependence of LINERs and Seyferts on host galaxies}

Let us briefly mention the partition between LINERs and Seyferts
among the Oxygen-excess objects as a function of host galaxy properties.
We define LINERs as {\sc [OII]}$_{AGN}>${\sc [OIII]}$_{AGN}$ as discussed in Section 3.2.
We remind the readers
again that the subscript {\it AGN} means that the luminosity due to
star formation is subtracted and the extinction correction described
in Section 3.1 has been applied.
\citet{kewley06} and \citet{schawinski07} studied the LINER/Seyfert partition
using BPT AGNs, but we revisit the subject with pure AGN luminosities.

%---------------------
\begin{figure}
  \begin{center}
    \FigureFile(80mm,80mm){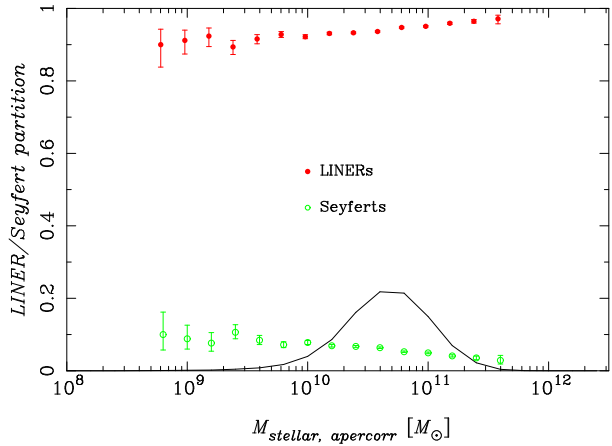}
  \end{center}
  \caption{
    Fractions of Seyferts and LINERs as a function of stellar mass.
    The filled and open circles show LINERs and Seyferts, respectively.
    The solid line shows the normalized distribution of all the Oxygen-excess galaxies.
  }
  \label{fig:agn_comp10}
\end{figure}
%---------------------

As in the last subsection, we start with stellar mass dependence
shown in Fig. \ref{fig:agn_comp10}.
It is striking that most Oxygen-excess objects are LINERs and
the LINER/Seyfert partition shows only very weak
dependence on stellar mass.  This is in contrast to the strong stellar
mass dependence of the AGN fraction discussed above.  AGNs prefer
massive galaxies, but once they become active, their activity
(LINER vs Seyfert) does not strongly depend on mass.
We will further discuss this 'mass switch' to activate AGNs in the next section.

%---------------------
\begin{figure*}
  \begin{center}
    \FigureFile(80mm,80mm){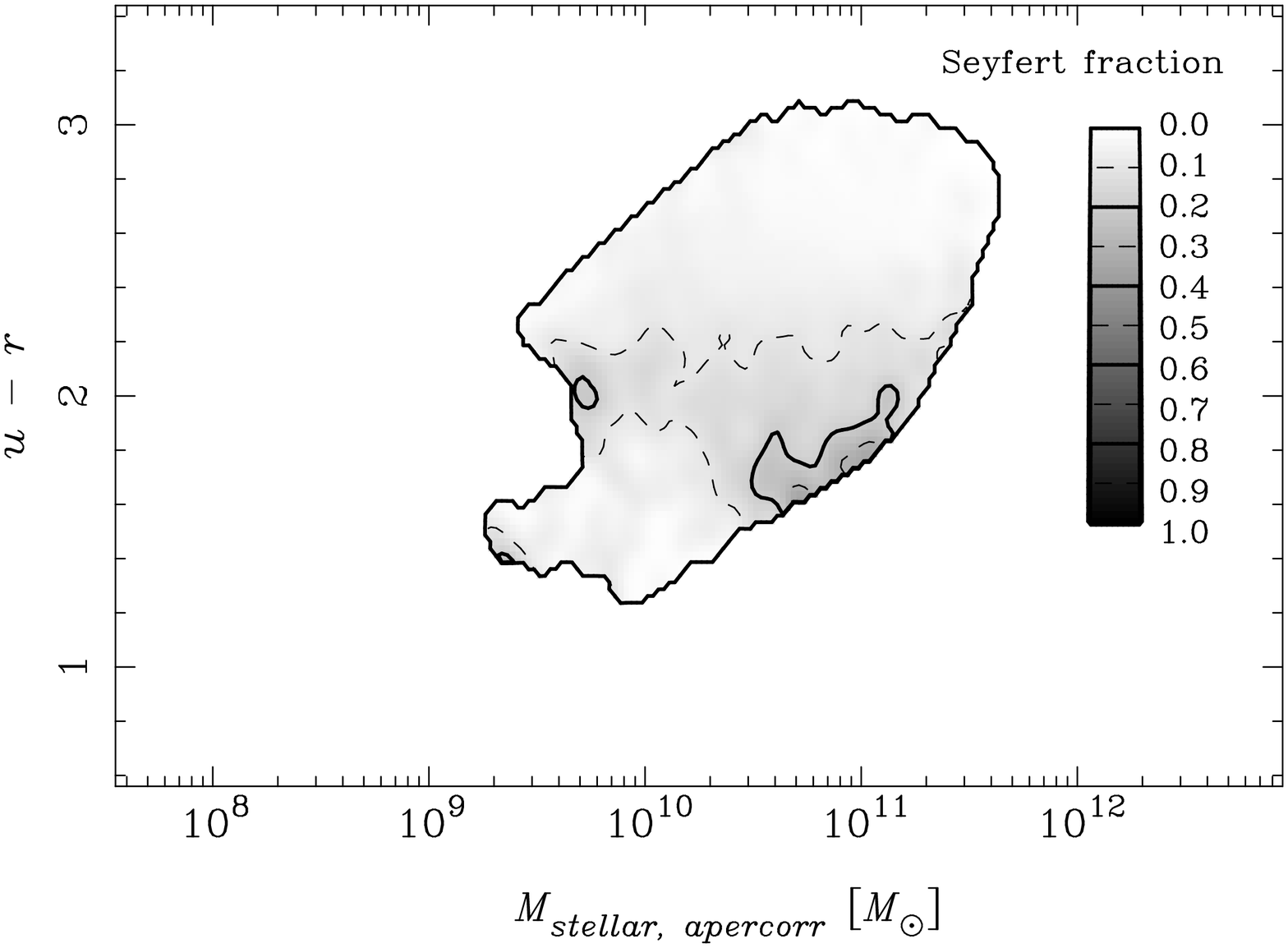}\hspace{0.5cm}
    \FigureFile(80mm,80mm){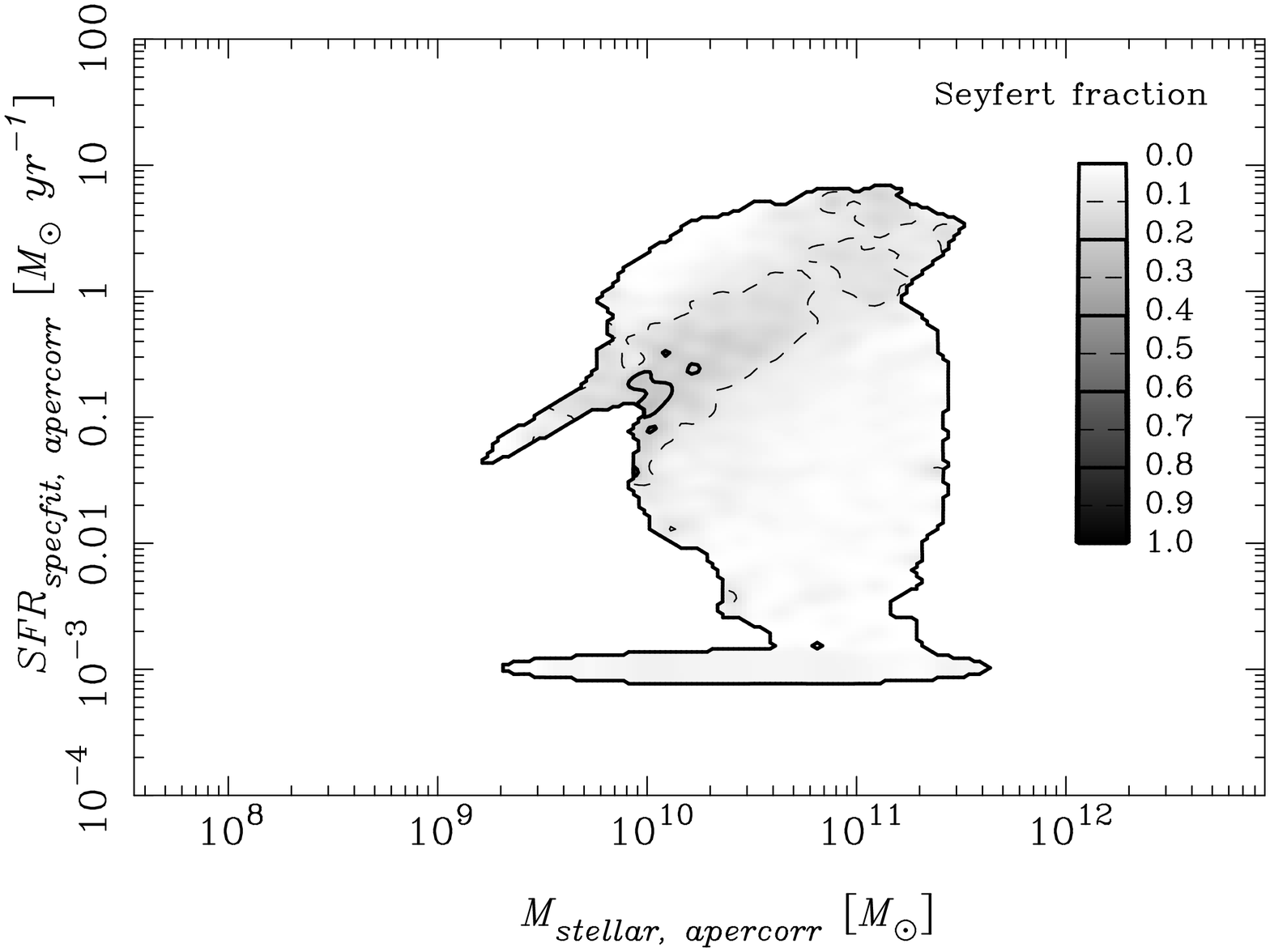}\\\vspace{0.5cm}
    \FigureFile(80mm,80mm){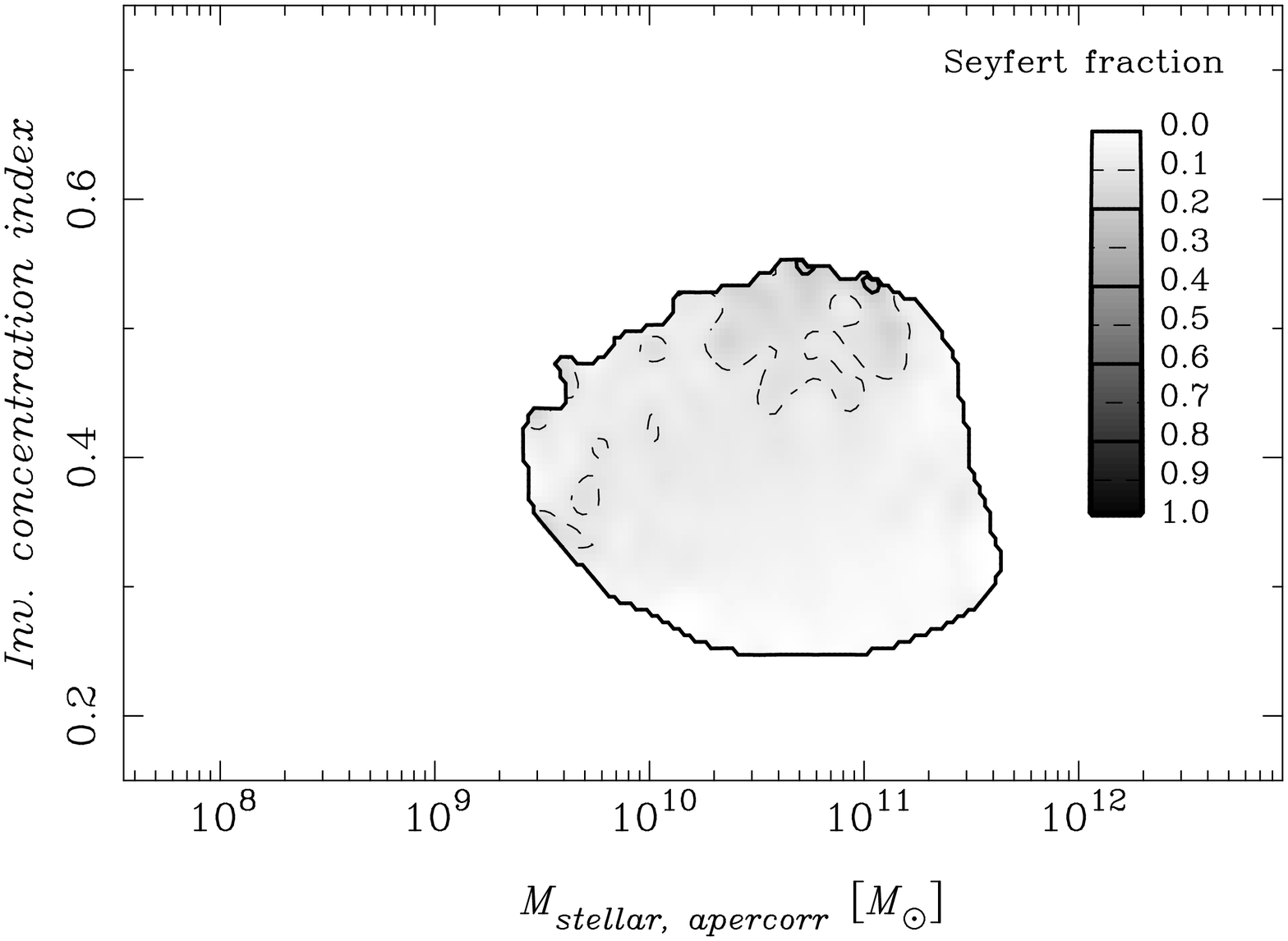}\hspace{0.5cm}
    \FigureFile(80mm,80mm){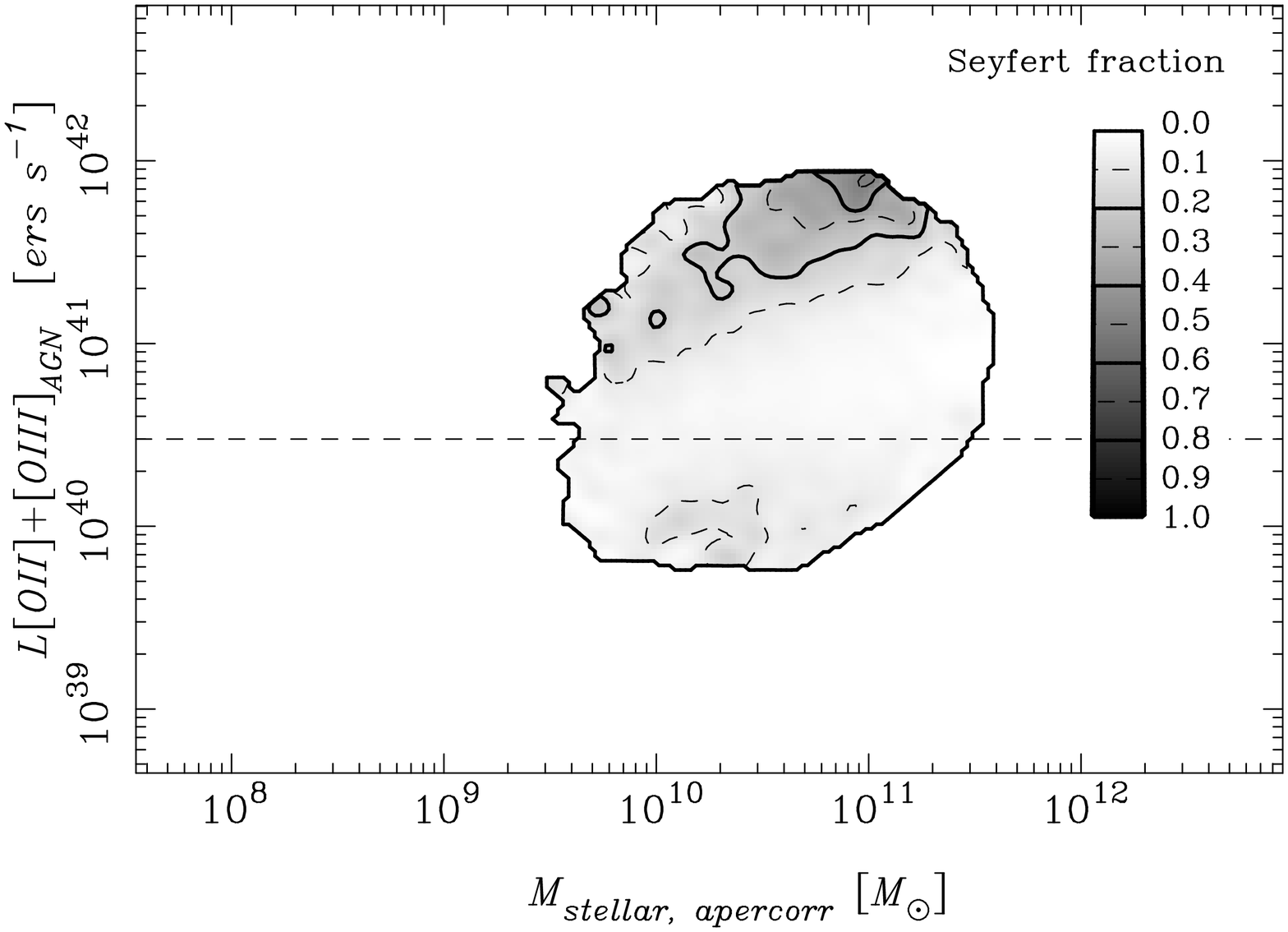}
  \end{center}
  \caption{
    Fraction of Seyfert galaxies ({\sc [oii]$_{AGN}$}$<${\sc [oiii]$_{AGN}$}) among
    all the Oxygen-excess objects.  The panels show the fraction as
    functions of stellar mass and rest-frame $u-r$ color (top-left), 
    SFR (top-right), inverse concentration index (bottom-left), and
    AGN power (bottom-right).
    We show our sensitivity limit of AGN power in the bottom-right
    panel as the dashed line.
    The contours are as in Fig. \ref{fig:agn_comp_color}.
  }
  \label{fig:agn_comp_seyfert}
\end{figure*}
%---------------------

Fig. \ref{fig:agn_comp_seyfert} shows the LINER/Seyfert partition
as functions of stellar mass, rest-frame $u-r$ color, SFR, morphology
and AGN power.
There is a weak trend that Seyferts increase in late-type, blue, star forming
galaxies.  Massive red, quiescent galaxies do not show Seyfert-like activities.
This is qualitatively consistent with \citet{kewley06}
who observed that typical age of stellar populations in Seyferts is
younger than that of LINERs.
It is also in agreement with \citet{schawinski07} that LINERs tend to be
in quiescent galaxies.
The LINER/Seyfert partition is apparently not strongly dependent on host galaxy properties, but
we find that the partition is a relatively strong function of the AGN power.
At $L_{[OII]+[OIII]_{AGN}}<10^{41}\rm\ erg\ s ^{-1}$, the Oxygen-excess objects
are mostly LINERs. 
The Seyfert fraction rapidly increases
towards high luminosity and Seyferts and LINERs almost equally exist
in the most powerful objects studied here. 
We note that, within the redshift range explored ($0.02<z<0.1$),
we are complete down to a luminosity of $L_{[OII]+[OIII]_{AGN}}=3\times10^{40}\rm\ erg\ s^{-1}$
for quiescent galaxies.
This number is derived from the lower boundary of galaxy distribution
on a $L_{[OII]+[OIII]_{AGN}}$ vs redshift plot.

To summarize, the partition is primarily a function
of AGN power and its dependence on host galaxy properties is relatively weak.
Seyferts seem to slightly prefer star forming galaxies, but this
correlation is likely due to  a correlation between SFR and AGN power presented below.

%------------------------------------------
\subsection{Correlations between AGN activity and host galaxy properties}

To further discuss the AGN activity, we introduce black hole mass.
There are a few ways to estimate black hole mass.
Here we take an empirical way and use the stellar velocity dispersion.
Bulge stellar
velocity dispersion is observed to exhibit a tight correlation with black
hole masses \citep{gebhardt00,ferrarese00,tremaine02,gueltekin09}.
We use the velocity dispersion estimates from the MPA/JHU catalog
\citep{tremonti04,kauffmann04} and derive the black hole masses
using the empirical relation from \citet{tremaine02}:

\begin{equation}
\log(M_{BH}/M_\odot)=8.13+4.02\log(\sigma/200),
\end{equation}

\noindent
where $\sigma$ is the stellar velocity dispersion in units of $\rm km\ s^{-1}$.
This calibration is consistent with a recent calibration \citep{gueltekin09}.
We should emphasize that our black hole mass estimates are nothing more than
a rough guess and are only of a statistical value as warned by previous studies
(e.g., \cite{kauffmann07}).
The SDSS measures the velocity dispersion within 3 arcsec fibers,
and hence it is measured over different physical sizes for galaxies at different redshifts.
A velocity component of disk will be a significant source of error
in our black hole mass estimates.  There is no straightforward way to
correct for such an effect and dedicated observations would be needed
for more accurate black hole mass estimates.
Note that we remove objects with $\sigma<70\rm km\ s^{-1}$ from the catalog
because of the limited SDSS spectral resolution of $R\sim2000$.

%---------------------
\begin{figure*}
  \begin{center}
    \FigureFile(80mm,80mm){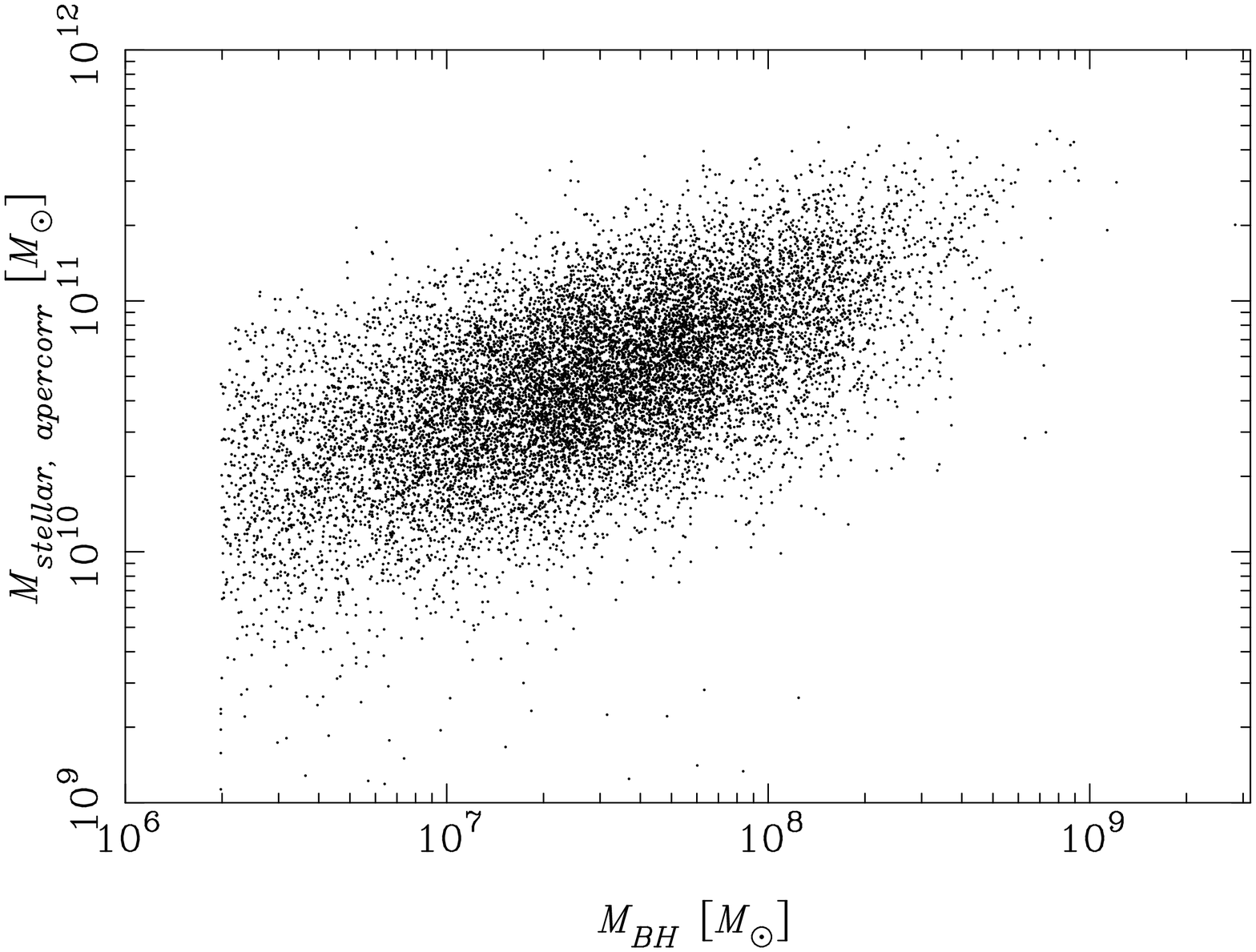}\hspace{0.5cm}
    \FigureFile(80mm,80mm){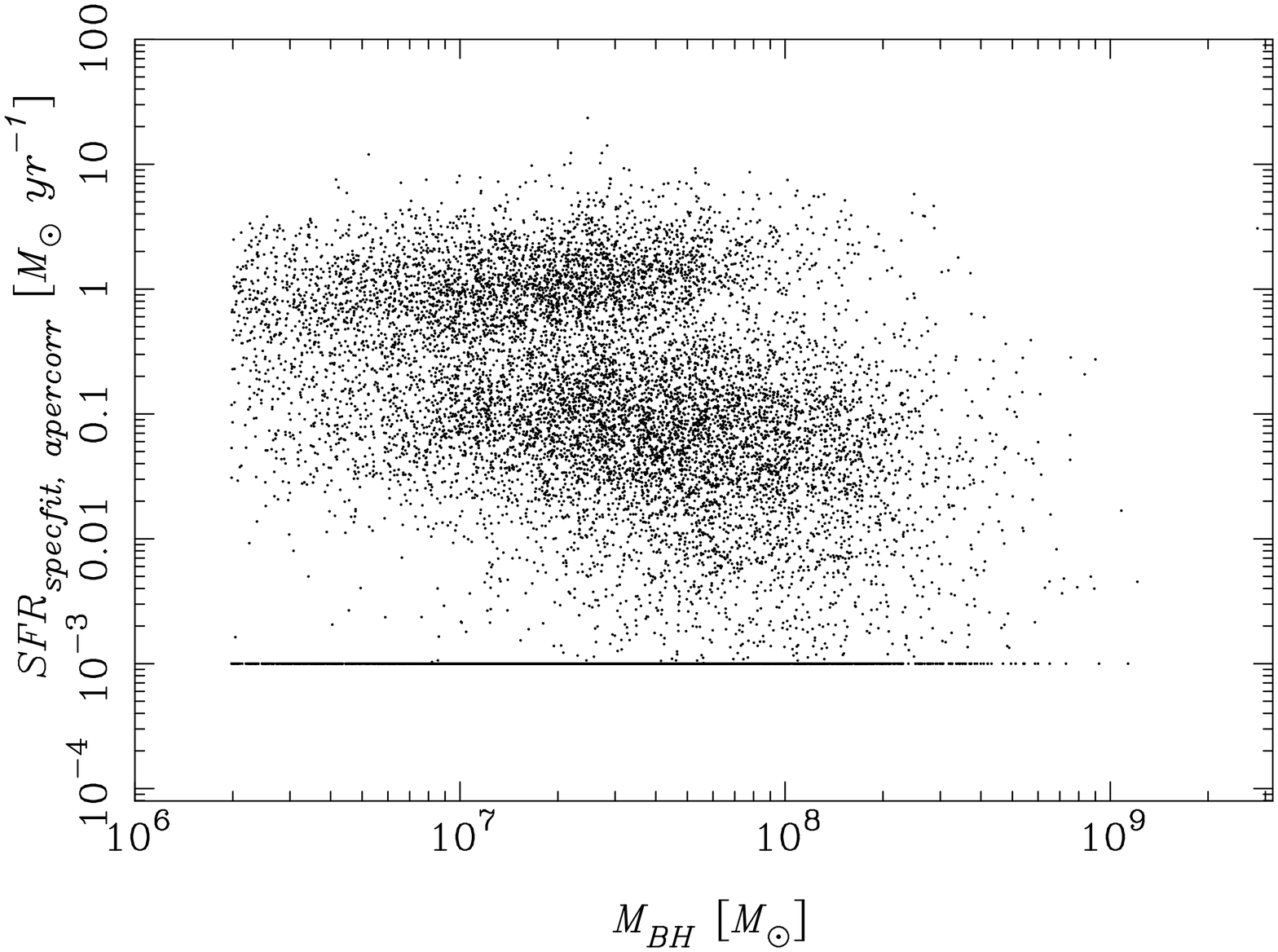}\\
    \FigureFile(80mm,80mm){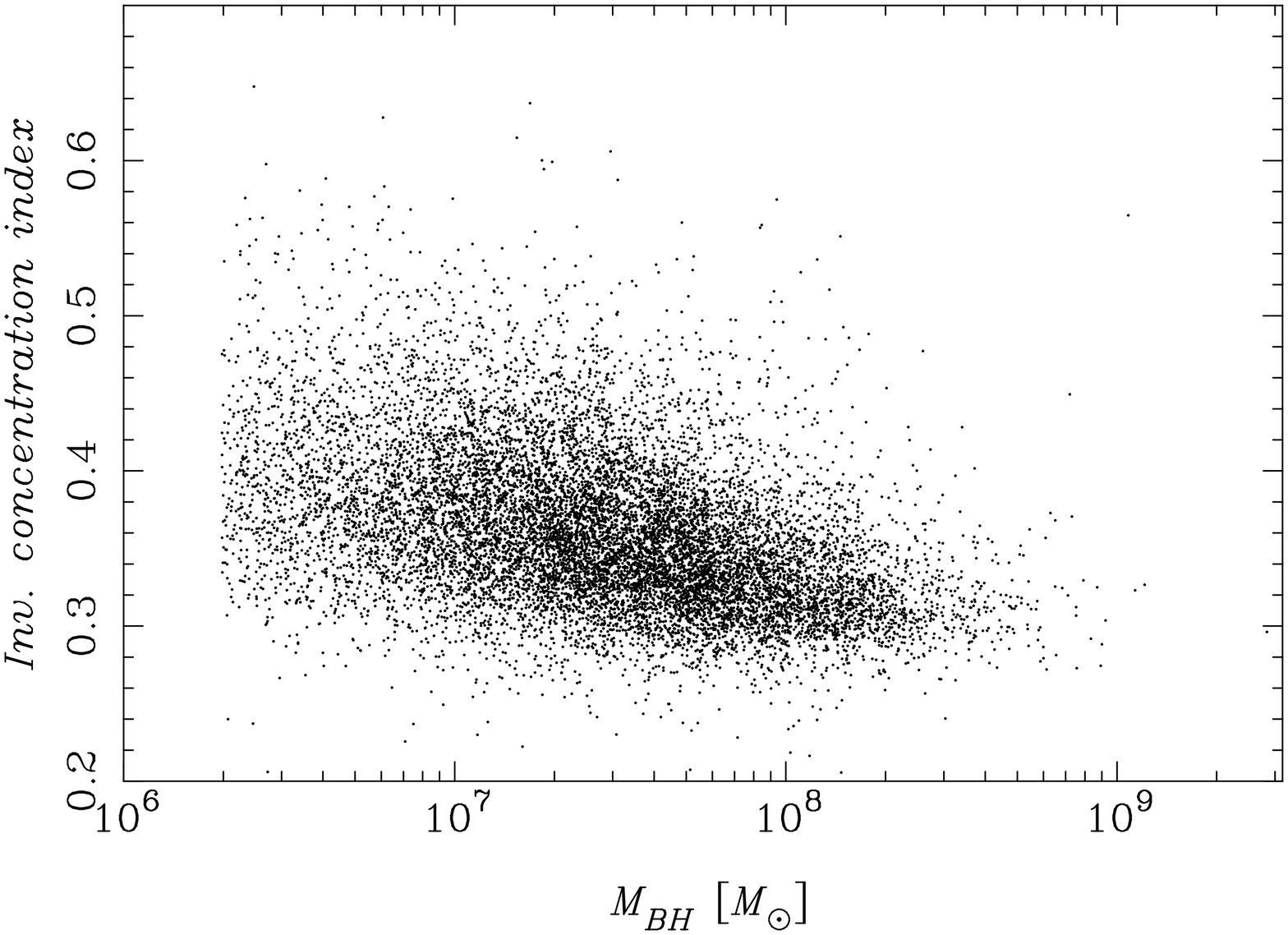}\hspace{0.5cm}
    \FigureFile(80mm,80mm){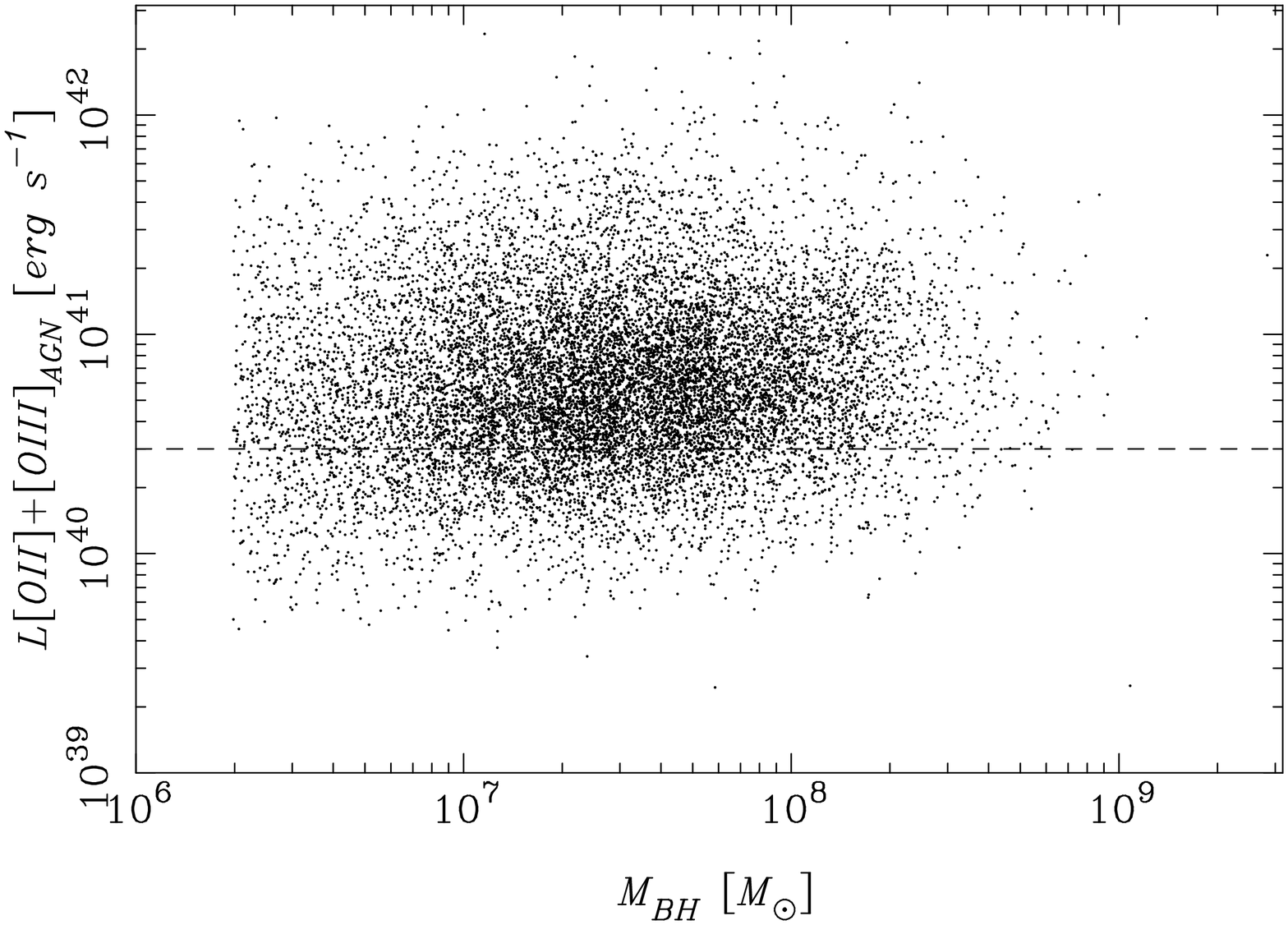}
  \end{center}
  \caption{
    Correlations between $M_{BH}$ and stellar mass (top-left),
    SFR (top-right), inverse concentration index (bottom-left),
    and  AGN power (bottom-right).
  }
  \label{fig:mbh_corr}
\end{figure*}
%---------------------

We show in Fig. \ref{fig:mbh_corr} correlations between the black hole mass
and stellar mass, SFRs, morphology, and AGN power.
Although not all the plots are entirely new and interesting,
we show them as a representative set of correlations between the primary parameters.
We make the following points from these plots.\\

\noindent
{\bf Stellar mass:}
Black hole mass increases with increasing stellar mass.
This correlation is expected given that more massive galaxies
show higher stellar velocity dispersions.
We obtain a Spearman's rank correlation coefficient of 0.58
with a null probability of $\ll10^{-5}$.
Most Oxygen-excess objects have $M_{BH}$ between $10^7$ and $10^8\rm\ M_\odot$.
This is similar to the black hole mass distribution of LINERs and
Seyferts in the Palomar survey \citep{ho09}.\\

\noindent
{\bf SFR:}
Given the clear positive correlation between stellar mass and black
hole mass, this plot is basically equivalent to a SFR vs stellar mass
plot.  Galaxies form two well-known sequences -- star forming and quiescent.
The bimodality is particularly clear in massive galaxies.\\

\noindent
{\bf Morphology:}
Most of the Oxygen-excess objects exhibit early-type morphology
with the inverse concentration index below 0.4 \citep{shimasaku01,strateva01}.
The index gradually increases towards low black hole mass,
but we are not sure if this is a real trend.  Lower mass objects appear
smaller on the sky and thus they are more strongly affected by 
the seeing effect, which results in increased inverse concentration index.
But, in any case, it is fair to say that most Oxygen-excess objects are
early-type galaxies.
\\

\noindent
{\bf AGN power:}
There seems to be a weak positive correlation between the AGN power and
black hole mass, although the scatter is very large.
A Spearman's correlation coefficient is 0.10, but the null hypothesis
is rejected with a $\ll10^{-5}$ probability.
Such a weak correlation is surprising given the strong stellar mass
dependence of the AGN fraction discussed in Section 4.1.
We recall that we did not observe a clear correlation between
hard X-ray luminosity and stellar mass in Fig. \ref{fig:oiioiii_xray5}.
A similar trend has been observed by other authors \citep{aird11,mullaney11}.
We will elaborate on the trend in Section 5.\\

%------------------------------------------
\subsection{The co-evolution of black holes and host galaxies}.

%---------------------
\begin{figure*}
  \begin{center}
    \FigureFile(120mm,120mm){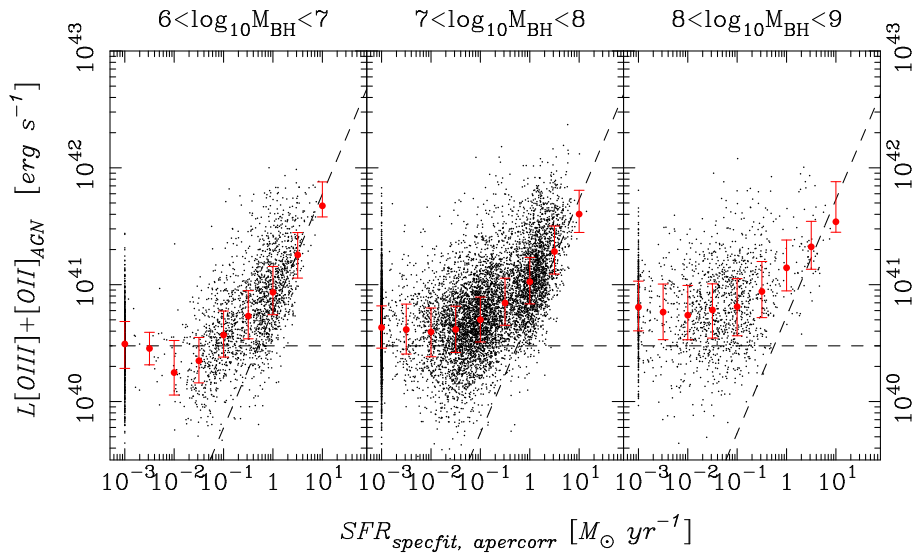}\\
  \end{center}
  \caption{
    AGN power plotted against redshift.
    The dots are Oxygen-excess objects and the large point with
    the error bar shows the median and quartile of $L[OII]+[OIII]_{AGN}$ of galaxies
    in each SFR bin.
    The horizontal dashed lines show detection limits and the slanted dashed lines
    are the selection function.
    We miss weak AGNs in actively star forming galaxies.
  }
  \label{fig:mbh_sfr_oiioiii1}
\end{figure*}
%---------------------

We have quantified the basic correlations between the primary parameters
and we now characterize relationships between activities and growth rates
of AGNs and those of host galaxies.
Fig. \ref{fig:mbh_sfr_oiioiii1} shows the correlation
between AGN power and host galaxy SFR.  This plot therefore
compares the activities of the host galaxies in terms of star formation
($\dot{M}_{stellar}$) and the accretion rate of
the central black hole ($\dot{M}_{BH}$).  
We recall that we have subtracted the star formation component from 
the {\sc [oii]+[oiii]} luminosity and corrected for extinction, thereby representing
intrinsic AGN emission.
The slanted dashed lines are our selection bias mentioned several times by now.
Because we require a significant $L_{AGN}$ with respect to $L_{SF}$ in Eq. 1,
we miss weak AGNs in actively star forming galaxies.
The dashed lines
are the limits where $L_{AGN}$ can be detected at $1.5\sigma$
($>1.5\sigma$ is the definition of Oxygen-excess) at a given $L_{SF}$
for galaxies with typical extinction.
Unlike BPT, the selection bias of the Oxygen-excess method is
understood fairly well.

The AGN power seems to increase with increasing SFRs.
But, the trend is driven by the incompleteness at
the bottom-right corner of the plots due to the selection bias.
The plot does not give us an average AGN power as a function
of SFR, and we can only discuss the upper envelope
of the AGN power distribution.
AGNs in low-SFR galaxies are typically of low-luminosity and
few powerful AGNs (several times $10^{41}\rm\ erg\ s^{-1}$, say)
are located in those galaxies.  The upper envelope of
the AGN power distribution extends to higher luminosity at higher SFRs and
powerful AGNs are often hosted by actively star forming galaxies.
This suggests that galaxies and the central black holes can become
active simultaneously.  This is a surprising phenomenon because galaxy star formation
and the central AGN activity occur on a very different physical scales
(10 kpc vs. 1 pc).

The AGN power is on average lowest at SFR$<0.1\rm \ M_\odot\ yr^{-1}$.
This weak AGN power in quiescent galaxies is again the point we have often
mentioned above.
Weak AGNs tend to reside in quiescent galaxies in our sample.
In other words, AGNs on the red sequence are low-luminosity AGNs.
These are the galaxies that BPT tends to miss;
as we have shown in Fig. \ref{fig:agn_comp_color}, the BPT gives a low AGN
rate on the red sequence, while we find a large AGN population there.
On the other hand, strong AGNs are located in actively star forming
galaxies and we find that they tend to fill in the green valley in line with
previous studies \citep{schawinski07,schawinski10}.

The AGN power - SFR relation observed in Fig.  \ref{fig:mbh_sfr_oiioiii1}
may provide a physical link between the down-sizing in galaxy
star formation (e.g., \cite{thomas05}) and the down-sizing in AGN
activities \citep{ueda03}.
More massive galaxies form at earlier times with shorter star formation
time scales.  The peak SFRs of galaxies are therefore higher for more massive galaxies.
Fig 10 of \citet{thomas05} qualitatively illustrates it.
If the AGN power is correlated with the SFRs as shown in
Fig. \ref{fig:mbh_sfr_oiioiii1} at all redshifts, then we expect to
observe a peak of space density of stronger AGNs at higher redshifts.
\citet{ueda03} actually observed that a peak of stronger AGN
activity is located at higher redshifts.
Perhaps the two down-sizing phenomena are linked by the co-evolution
of black holes and host galaxies.

%---------------------
\begin{figure*}
  \begin{center}
    \FigureFile(120mm,80mm){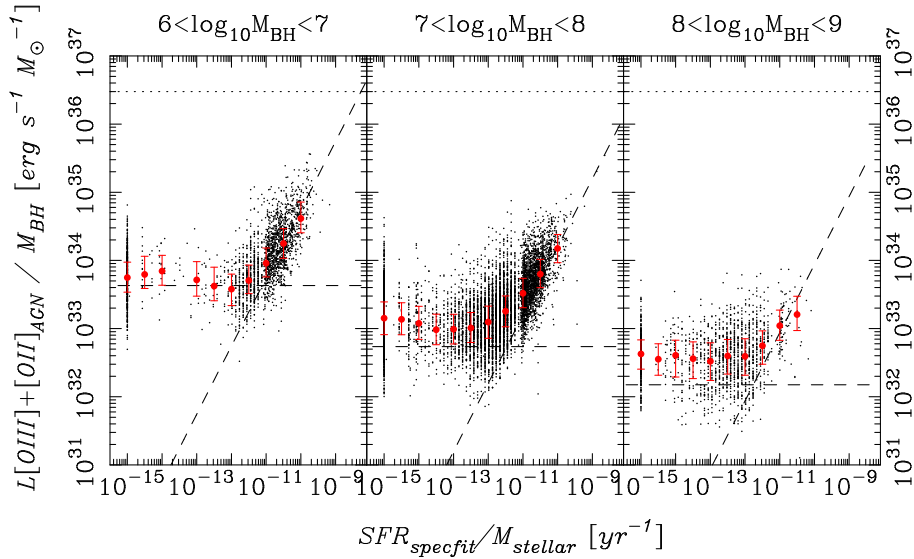}\\
  \end{center}
  \caption{
    $L_{[OII]+[OIII]_{AGN}}/M_{BH}$ plotted against SFR/$M_{stellar}$.
    The dashed lines are our detection limit and selection function.
    The dotted line is an approximate Eddington limit.
    The galaxies are distributed in vertical stripes due to the discrete model grids.
  }
  \label{fig:mbh_sfr_oiioiii2}
\end{figure*}
%---------------------

To obtain deeper insights into the relationship between the AGNs and host galaxies,
we compare black hole growth rate ($L_{[OII]+[OIII]_{AGN}}/M_{BH}$, which is proportional to
$\dot{M_{BH}}/M_{BH}$), and host galaxy growth rate ($SFR/M_{stellar}$,
which is $\dot{M}_{stellar}/M_{stellar}$) in Fig. \ref{fig:mbh_sfr_oiioiii2}.
$L_{[OII]+[OIII]_{AGN}}/M_{BH}$ also serves as an indicator of the Eddington ratio.
We observe a tight correlation between H$\alpha$ and {\sc [oii]+[oiii]}
($L_{[OII]+[OIII]_{AGN}}\sim4.5 L_{H\alpha_{AGN}}$) and we obtain a bolometric correction
factor of 50 to $L_{[OII]+[OIII]_{AGN}}$ assuming $L_{bol}\sim220L_{H\alpha}$ from \citet{ho08}.
We find that most of the Oxygen-excess objects typically have an Eddington
ratio between $10^{-4}$ and $10^{-2}$.
We note that a bolometric correction to convert an emission line luminosity into
bolometric luminosity is highly uncertain and probably not a constant correction \citep{ho08}.
Therefore, the Eddington ratio should not be over-interpreted.

We are again  limited by the sensitivity limit and selection bias, but galaxies with
lower mass black holes (i.e., lower mass galaxies) tend to show higher
black hole growth rates and galaxy growth rates.  
In any black hole mass bin, the black hole growth rate is low
at low SFR$/\rm M_{stellar}$ and few objects with high $L_{[OII]+[OIII]}\rm /M_{BH}$
are observed.  The upper envelope of the black hole growth rate
increases with increasing host galaxy growth rate and
rapidly growing black holes are always hosted by rapidly
growing galaxies.
This clearly represents the co-evolution of the super-massive black holes
and their host galaxies.
Again, this is surprising because these two growths occur at very different
physical scales.  The black hole growth occurs on a 1 pc scale (or less),
while the galaxy growth occurs on a 10 kpc scale.
But, our result shows that black holes grow in growing galaxies.

There are many recent speculations on energy feedback from AGNs to
suppress star formation activities of the host galaxies.
Do we observe any hint of AGN feedback in action in our data?
The positive correlation in the upper envelope of the AGN power and
host galaxy SFR suggests that strong AGNs do not appear to suppress on-going
star formation.
If strong AGNs could sharply suppress star formation,
we would  not have observed the positive correlation between AGN power and
host galaxy SFRs.
However, there may be a time delay between AGN activities and the suppression
of SFRs.  Also, the apparent lack of suppressed SFRs
may possibly be because we do not probe very strong AGNs such as QSOs,
in which star formation may be suppressed \citep{ho05,kim06}.

On the other hand, the observed large fraction of AGNs in quiescent galaxies
may give a support to the so-called 'radio mode' feedback \citep{croton06,bower06},
which suppresses gas cooling for further star formation and keeps red galaxies red.
Our data, however, do not tell us whether these AGNs are actually suppressing
the gas cooling and they have any direct effect on the host galaxies.
We cannot make anything more than a speculation here, and
it would be fair to say that we do not observe any convincing evidence of
AGN feedback in action and we do not observe any evidence against it.

Some of the results in this subsection are already mentioned in previous studies
using colors as a proxy for star formation rate.
But, we have characterized them with physical parameters such as
host galaxy SFRs.  We shall also emphasize again that
we have studied intrinsic AGN emission thanks to the unique feature of
the Oxygen-excess method taking into account the selection bias.
We have put the previous results on the quantitative ground and
it is now clear that growing black holes are often hosted by growing galaxies,
which represents the co-evolution of black holes and the host galaxies.

%--------------------------------------------------------------------------
\section{Summary and Discussion}

In the last section, we have studied correlations between host galaxy properties
and AGN activities. Among the numerous plots presented there, our primary
findings can be summarized as follows.

\begin{enumerate}
\item The AGN fraction increases with increasing stellar mass of the host galaxies.
Most of the AGNs are located in $>10^{10}\rm\ M_\odot$ galaxies.
\item The AGN fraction is higher in red early-type galaxies than in blue late-type galaxies.
But, this is at least partly driven by the fact that the Oxygen-excess method tends to miss low-luminosity AGNs
in actively star forming galaxies.
\item Most of the identified AGNs are LINERs.  The LINER/Seyfert partition is
primarily a function of AGN power and its dependence on host galaxy properties is weak.
Seyfert-like AGNs with {\sc [oii]$_{AGN}$}$<${\sc [oiii]$_{AGN}$}
appear only in powerful AGNs.
\item AGN power does not strongly correlate with black hole mass (or stellar mass of the host).
\item Powerful AGNs are located in actively star forming galaxies.
\item Rapidly growing black holes are hosted by rapidly growing galaxies.
\item We observe no direct evidence of the AGN feedback in action.
\end{enumerate}

We find the 1st and 4th points above very interesting.
The first point is shown in Fig. \ref{fig:agn_comp1} and
the 4th point is in Fig. \ref{fig:oiioiii_xray5} and the bottom-right panel of
Fig. \ref{fig:mbh_corr}.
Mass is the primary parameter to
characterize the fraction of AGNs.  The AGN fraction is very small below
$10^{10}\rm\ M_\odot$, but it increases rapidly above that mass.
One might suspect that the lower AGN fraction at lower mass is because
lower mass black holes show weaker activities
and they go undetected.
However, the 4th point above suggests that is not the case.
The correlation between AGN power and mass is very weak and
it is unlikely that such a weak correlation
can explain the observed strong mass dependence of the AGN fraction.
Recently, \citet{mullaney11} reached the same conclusion.
Furthermore, lower-mass galaxies are located at lower redshifts
due to the nature of a flux limited survey.
The detection limit on $L_{[OII]+[OIII]_{AGN}}$ shown by the dashed lines in the figures
is a limit at which we can detect it at all redshift ($0.02<z<0.10$) and our sensitivity
is actually better for low-mass galaxies which we can observe
only at low redshifts.
The observed lack of low-mass AGNs is not due to the absence of bulge either.
As shown in Fig. \ref{fig:agn_comp_cin}, the AGN fraction is very low at
low mass regardless of the morphology.
While very low-mass AGNs do exist (e.g., NGC4395 has
$M_{BH}\sim4\times10^5\rm M_\odot$; \cite{peterson05}; see also the work by \cite{barth08}),
the observed strong mass dependence suggests
that such low-mass AGNs are rare.
It seems that mass is like a 'switch' to activate super-massive black holes.

There are at least two very naive interpretations of this absence of low-mass AGNs.
One is that super-massive black holes exist in all galaxies, but low-mass galaxies cannot
transport material to the center and feed super-massive black holes for some reason.
That is, the majority of the black holes are inactive and thus are not detected.
In this scenario, a process to trigger AGN activity is unlikely related to
the presence of bulge because the AGN fraction is nearly zero in low-mass
galaxies regardless of their morphology (Fig. \ref{fig:agn_comp_cin}).
Also, the presence/lack of a transportation mechanism must be
largely independent of the accretion rate as we observed no strong
correlation between the AGN power and black hole mass (Figs. \ref{fig:oiioiii_xray5}
and \ref{fig:mbh_corr}).
Namely, a mass transportation mechanism is established at a higher rate
in more massive galaxies, but the transportation rate should
be largely independent of mass.
Although such a physical process is somewhat difficult to imagine,
this is a possible interpretation of the result.

The other very naive interpretation is that the majority of low-mass galaxies do not host super-massive black holes.
This is perhaps an extreme idea, but it does explain the observed trend
because the presence/absence of super-massive black holes works just like a switch.
If a galaxy hosts a super-massive black hole, it can be active if the host can feed
it.  If a galaxy does not harbor a super-massive black hole, it can never
be active regardless of the host galaxy properties.  In this scenario,
mass of the host galaxies is a probability function of the presence
of super-massive black holes.  The form of the probability function is such that
more massive galaxies are more likely to host super-massive black holes and
the probability is very small at  $<10^{10}\rm\ M_\odot$.
In fact, the super-massive black hole occupation fraction is apparently
small in low-mass galaxies (e.g., M33 and NGC205 do not host a
super-massive black hole; \cite{gebhardt01,valluri05}).
This provides supportive evidence of the decreasing probability of hosting
super-massive black holes with decreasing the host galaxy mass.

\citet{mullaney11} observed that the AGN activity is nearly independent
of host galaxy stellar mass at $0<z<3$.
They also observed that AGNs are almost exclusively hosted by $>10^{10}\rm M_\odot$
galaxies.  It seems that the trends that we observe at $z=0$ holds out to $z=3$.
This might suggest that a primary epoch of the formation of super-massive black holes
is located at very high redshift.
Dark matter halos that collapse at high redshifts are among the most massive ones today
and therefore the dominance of AGNs in massive galaxies is naturally
explained if the formation of super-massive black holes peaked at high redshifts.
The absence of low-mass AGNs might in turn imply that the super-massive black hole
formation is not very efficient at low redshift.
Perhaps the mass dependence of the AGN fraction reflects the formation
history of the super-massive black holes.

Our data do not allow us to conclude which of the two scenarios
discussed above is more likely.
Also, there may well be other ways to interpret our result.
Before we try to make any further speculations, we have to
make an attempt to confirm that an AGN fraction in low-mass galaxies is
actually very low compared to that in high-mass galaxies.  An SDSS
fiber subtends 3 arcsec on the sky, which corresponds to $\sim3$ kpc
at $z=0.05$.  AGN emission, if any, is therefore strongly contaminated
by host galaxies. 
Also, our black hole mass estimates are probably not very accurate.
\citet{greene07} argued that the black hole mass function
decreases at low mass.  Our result may not be directly comparable
to theirs because their result is based on objects with broad
H$\alpha$ emission, while ours is based on narrow line AGNs.
But, the observed decline in the black hole mass function at
low-mass end is qualitatively consistent with ours.
Still, it would be necessarily to carry out a dedicated
observation to sample the nuclear spectra of nearby low-mass galaxies
to achieve a better sensitivity to identify AGNs.
If the absence of AGNs in low-mass galaxies is confirmed in such
a survey, then we will be in a position to further discuss the physical origin
of such a trend.

\vspace{0.5cm}
We thank John Silverman for extensive discussions, Luis Ho, and Yoshihiro Ueda
for useful conversations,
Yen-Ting Lin for helpful comments, and Naoki Yasuda for providing the computing environment.
This work was supported by World Premier International Research Center Initiative
(WPI Initiative), MEXT, Japan and also in part by KAKENHI No. 23740144.
This research has made use of data obtained from
the Chandra Source Catalog, provided by the Chandra X-ray Center (CXC)
as part of the Chandra Data Archive.
We thank the anonymous referee for his/her useful comments, which 
helped improve the paper.

Funding for the Sloan Digital Sky Survey (SDSS) and SDSS-II
has been provided by the Alfred P. Sloan Foundation, the
Participating Institutions, the National Science Foundation,
the U.S. Department of Energy, the National Aeronautics and
Space Administration, the Japanese Monbukagakusho, and
the Max Planck Society, and the Higher Education Funding
Council for England. The SDSS Web site is http://www.sdss.org/.

The SDSS is managed by the Astrophysical Research Consortium
(ARC) for the Participating Institutions. The Participating
Institutions are the American Museum of Natural History,
Astrophysical Institute Potsdam, University of Basel,
University of Cambridge, Case Western Reserve University,
The University of Chicago, Drexel University, Fermilab,
the Institute for Advanced Study, the Japan Participation Group,
The Johns Hopkins University, the Joint Institute for Nuclear
Astrophysics, the Kavli Institute for Particle Astrophysics
and Cosmology, the Korean Scientist Group, the Chinese Academy 
of Sciences (LAMOST), Los Alamos National Laboratory,
the Max-Planck-Institute for Astronomy (MPIA), the 
Max-Planck-Institute for Astrophysics (MPA), New Mexico
State University, Ohio State University, University of Pittsburgh,
University of Portsmouth, Princeton University, the United 
States Naval Observatory, and the University of Washington.

%-------------------------------------------------------------------------------
%-------------------------------------------------------------------------------

\end{document}